\documentclass[manuscript,acmsmall]{acmart}

\usepackage{algorithmicx}
\usepackage{xspace}
\usepackage{multirow}
\usepackage{algorithm}
\usepackage{algpseudocode}
\usepackage{bm}
\usepackage{subcaption}

\newcommand{\app}{PromCopilot\xspace}
\AtBeginDocument{%
  }

\begin{document}

\title[PromCopilot]{PromCopilot: Simplifying Prometheus Metric Querying in Cloud Native Online Service Systems via Large Language Models}


\author{Chenxi Zhang}
\email{zhangchenxi@xidian.edu.cn}
\affiliation{%
  \institution{Xidian University}
  \city{Xi'an}
  \country{China}
}

\author{Bicheng Zhang}
\email{bczhang22@m.fudan.edu.cn}
\affiliation{%
  \institution{Fudan University}
  \city{Shanghai}
  \country{China}
}

\author{Dingyu Yang}
\email{yangdingyu@zju.edu.cn}
\affiliation{
  \institution{The State Key Laboratory of Blockchain and Data Security, Zhejiang University; Hangzhou High-Tech Zone (Binjiang) Institute of Blockchain and Data Security}
  \country{China}
}

\author{Xin Peng}
\authornote{Xin Peng is the corresponding author.}
\email{pengxin@fudan.edu.cn}
\affiliation{%
  \institution{Fudan University}
  \city{Shanghai}
  \country{China}
}

\author{Miao Chen}
\email{22210240006@m.fudan.edu.cn}
\affiliation{%
  \institution{Fudan University}
  \city{Shanghai}
  \country{China}
}

\author{Senyu Xie}
\email{24210240059@m.fudan.edu.cn}
\affiliation{%
  \institution{Fudan University}
  \city{Shanghai}
  \country{China}
}

\author{Gang Chen}
\email{cg@zju.edu.cn}
\affiliation{%
  \institution{The State Key Laboratory of Blockchain and Data Security, Zhejiang University; Hangzhou High-Tech Zone (Binjiang) Institute of Blockchain and Data Security}
  \country{China}
}

\author{Wei Bi}
\email{biwei.bw@alibaba-inc.com}
\affiliation{%
  \institution{Alibaba Group}
  \country{China}
}

\author{Wei Li}
\email{jianhao@taobao.com}
\affiliation{%
  \institution{Alibaba Group}
  \country{China}
}








\renewcommand{\shortauthors}{Zhang et al.}

\begin{abstract}
With the increasing complexity of modern online service systems, understanding the state and behavior of the systems is essential for ensuring their reliability and stability.
Therefore, metric monitoring systems are widely used and become an important infrastructure in online service systems.
Engineers usually interact with metrics data by manually writing domain-specific language (DSL) queries to achieve various analysis objectives.
However, writing these queries can be challenging and time-consuming, as it requires engineers to have high programming skills and understand the context of the system.
In this paper, we focus on PromQL, which is the metric query DSL provided by the widely used metric monitoring system Prometheus.
We aim to simplify metrics querying by enabling engineers to interact with metrics data in Prometheus through natural language, and we call this task text-to-PromQL.
Building upon the insight, this paper proposes \app, a Large Language Model-based text-to-PromQL framework.
\app first uses a knowledge graph to describe the complex context of a cloud native online service system.
Then, through the synergistic reasoning of LLMs and the knowledge graph, \app transforms engineers' natural language questions into PromQL queries.
To evaluate \app, we manually construct the first text-to-PromQL benchmark dataset which contains 280 metric query questions.
The experiment results show that \app is effective in text-to-PromQL.
When using GPT-4 as the backbone LLM, \app achieves an accuracy of 69.1\% in translating natural language questions to PromQL queries when using.
To the best of our knowledge, this paper is the first study of text-to-PromQL, and \app pioneered the DSL generation framework for metric querying and analysis.
\end{abstract}



\maketitle

\section{Introduction}
\label{section:introduction}

In recent years, online service systems represented by search engines, online shops, and social networks have become increasingly popular and important in our daily life.
Moreover, with the development of technologies such as cloud computing, containers, and microservices, these online service systems have become increasingly complex and large~\cite{trainticket-tse}.
Industrial online service systems may contain tens to hundreds of nodes, hundreds to thousands of services, and deployed on multiple data centers.
For such a complex system, it is crucial to observe and understand its state and behavior~\cite{ob-indstr-survey, sre}.

Therefore, metric monitoring systems are widely used in online service systems and have become an important infrastructure~\cite{sre,ob-indstr-survey}.
Metrics are measurements of resource usage or behavior that can be observed and collected throughout the system.
Monitoring systems enable engineers to collect, store, query, and visualize metrics in real-time.
Engineers usually set multiple types of metrics for each type of component in the system (e.g., nodes, service instances, services, APIs, etc.)~\cite{sre}, e.g., CPU and memory usage of nodes, queries per second (QPS) and response latency of APIs.
Modern online service systems usually includes a large number of metrics, which is because the whole system usually collect dozens to hundreds types of metrics, and each type of metric will generate many metric instances according to their associated system components.
Engineers usually achieve fault diagnosis, performance optimization, and other objectives by analyzing these metrics.

To help engineers interact with the massive metric data, most monitoring systems provide query interfaces based on domain-specific languages (DSL).
Engineers can query and analyze metric data by manually writing DSL queries.
For example, Prometheus provides PromQL~\cite{promql}, InfluxDB provides Flux~\cite{InfluxDB}, and Microsoft Azure provides Kusto~\cite{azure}.
In this paper, we focus on Prometheus~\cite{prometheus}, one of the most widely used metric monitoring systems in modern online service systems, and it has been recognized as the de facto standard in the field of cloud-native-based online service system monitoring~\cite{Prom_ali,Prom_azure,Prom_report}.
The query language provided by Prometheus is called PromQL~\cite{promql} (Prometheus Query Language), which enables engineers to query and analyze the metric data in Prometheus.


Manually writing PromQL queries is usually a challenging task.
It requires the engineer to understand not only the syntax of PromQL but also the context of the target system related to the query.
Specifically, first, a PromQL query can be very complex in real online service systems, which consist of multiple metric names, metric labels, metric values, operators, functions, etc.
It requires engineers have an depth understanding of the systax of PromQL and extensive experience in writing PromQL queries, which brings a large learning cost.
Second, the large scale and complexity of modern online service systems make it difficult for engineers to understand the full context of the system.
This makes engineers may spend a lot of time to search the system context knowledge when writing PromQL queries.
As the example shown in Figure~\ref{fig:motivation example}, the engineer wants to write a PromQL query to find which node has the most available memory among the nodes where the order service is deployed.
In order to write this PromQL query correctly, the engineer needs to know the correct metric name, the correct metric label, and where the order service (which may have dozens of service instances deployed on different nodes) is deployed.
Engineers usually acquire the above knowledge by searching through the various platforms and tools in the system, which is usually time consuming.

In order to simplify the process of metrics querying, in this paper we wanted to design an approach that enables engineers to simply interact with metric data in Prometheus through natural language.
Specifically, we would like engineers just to provide query requirements described in natural language, and then the approach can generate the corresponding PromQL queries.
We named this task as text-to-PromQL.
This kind of approach has been proven efficient in improving the productivity of engineers, such as the widely used techniques: code generation~\cite{llm_nl2code,csur_code_survey} and text-to-SQL~\cite{text-to-sql-survey1,text-to-sql-survey2}.
However, as shown in the example in Figure~\ref{fig:motivation example}, metrics queries are tightly tied to a specific system context, making it significantly different from the the usage scenarios of code and SQL.
Therefore, these existing techniques are difficult to apply to text-to-PromQL tasks.
A recent study~\cite{xpert} uses LLMs and few-shot learning to recommend metric queries related to an incident in cloud systems.
However, this approach only recommends metric queries based on specific incidents and cannot generate metrics queries on demand, which limits its usage scenarios.

To address the preceding challenges, this paper presents \app, an LLM-based text-to-PromQL framework.
\app can convert users' natural language questions into PromQL queries that match the context of a specific system, thus simplifying the process of metrics querying and analysis.
The main idea of \app is to describe the complex context of a specific system (e.g., metric information, component dependencies) through a knowledge graph, and then transform natural language questions into PromQL queries through the synergistic reasoning of knowledge graph and LLMs.
Specifically, \app first parses the input natural language question to understand the user intent and obtains the key information in the question.
Then it retrieves metric knowledge and system component knowledge related to the question on the knowledge graph. 
Finally, the retrieved knowledge is fed to the LLMs as context to help LLMs generate accurate PromQL queries.

To evaluate the effectiveness of \app, we manually construct the first text-to-PromQL benchmark dataset based an open-source microservice benchmark system.
Then we conduct a series of experimental studies on this dataset.
When using GPT-4-Turbo as the backbone LLM, \app achieves 90.3\% and 90.8\% in terms of recall for metrics retrieval and reasoning paths retrieval, and it finally achieves an accuracy of 69.1\% in translating natural language questions to PromQL queries.
The results demonstrate the effectiveness of \app in text-to-PromQL.
To the best of our knowledge, this paper is the first study for text-to-PromQL.

In summary, this paper makes the following main contributions:

\begin{itemize}
    \item We propose \app, a text-to-PromQL framework based on synergistic reasoning of knowledge graph and LLMs.
    \item We manually construct the first text-to-PromQL benchmark dataset, which is publicly available at~\cite{PromCopilot}.
    \item We conduct a series of experimental studies to validate the effectiveness of \app.
\end{itemize}

\section{Background and Motivation}
\label{section:background}

In this section, we first introduce the background about Prometheus and PromQL, then motivate our work with an example.

\subsection{Background}

Metric monitoring systems have become an important infrastructure for modern online service systems for understanding system behaviors and states~\cite{ob-indstr-survey,sre}.
Metric monitoring systems usually collect various metrics to observe the system state, and provide visualization and query functions to help engineers use the metrics data.
In general, metrics are numerical measurements of the system state over a period of time, commonly used metrics include CPU usage, requests per second, request error rate, etc.
Modern online service systems usually includes a large number of metrics, which is because the whole system usually collect dozens to hundreds types of metrics, and each type of metric will generate many metric instances according to their associated system components.
By observing and analyzing metrics data, engineers can detect system failures, identify root causes, and make operational decisions.

In recent years, Prometheus\cite{prometheus} has become the de facto standard in the field of cloud-native-based online service systems monitoring, because of its usability, scalability, and ecosystem.
Prometheus is an open-source monitoring and alerting toolkit, which is also one of the core projects of Cloud Native Computing Foundation\cite{cncf}.
Prometheus provides a multi-dimensional metrics data model, an easy-to-use metrics query language, an efficient time series database, and the ability to integrate with a wide range of third-party systems and tools. 
Many organizations have used Prometheus to build their monitoring systems, and cloud providers such as AWS\cite{aws}, Azure\cite{azure}, and Alibaba Cloud\cite{alibaba_cloud} also offer Prometheus cloud services.

\begin{figure}[t]
	\centering
	\includegraphics[width=0.8\linewidth ]{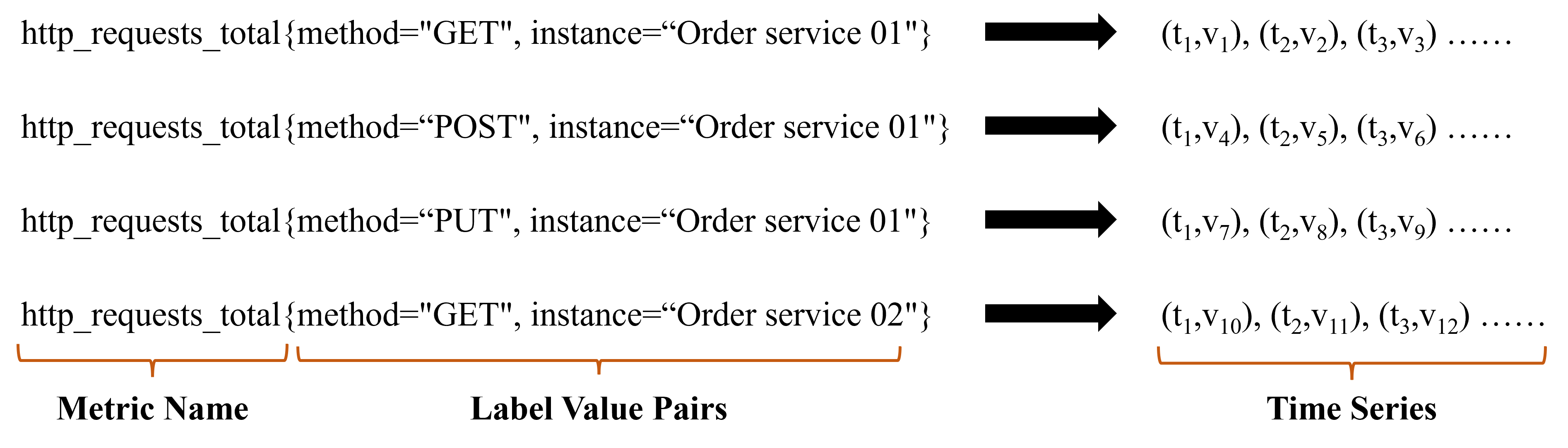}
	\caption{An Example of the Metric Data in Prometheus}
	\label{fig:datamodel example}
\end{figure}

The data model is the foundation of the metric monitoring system. 
In Prometheus's data model, a metric record will contain three parts of information: metric name, labels, and sample.
The metric name is usually a string that uniquely identifies a metric and briefly describes its monitoring object.
Labels are sets of label value pairs that help categorize and differentiate the different dimensions of a metric.
A sample contains a timestamp $t$ and a value $v$, which records the value of a metric under a set of label value pairs at a specific time.
Given a metric name and a set of label value pairs, a subset of samples can be filtered out.
When these samples are sorted by time, one or more time series can be obtained.
Figure~\ref{fig:datamodel example} shows an example of time series derived from a metric.
$http\_requests\_total$ is the name of the metric, which monitors the total number of HTTP requests.
The metric has two labels: $method$ and $instance$.
$method$ represents the HTTP request method and $instance$ represents the service instance that receives the request.
As shown in the figure, when we set different values for $method$ and $instance$ (e.g., $method$ as GET and $instance$ as Order service 01), we can obtain the corresponding time series consisting of sorted samples.

To help engineers easily work with metric data, metric monitoring systems usually provide query interfaces based on domain-specific languages(DSL).
Prometheus has implemented its own query language called PromQL\cite{promql} (Prometheus Query Language).
PromQL is a domain-specific language built upon Go, which is similar to SQL for managing databases, GraphQL for query graph databases, etc.
Using PromQL engineers can query and extract valuable insights from the time series in Prometheus.
PromQL supports string, scalar, range vector, and instant vector data types, where range Vector and Instant Vector are specifically designed by Prometheus for retrieving time series.
PromQL also provides many binary and aggregation operators and several functions to help engineers operate the data, such as add, subtract, sum, group, absolute values, etc.
For example, query $sum(rate(http\_requests\_total[5m]))$ calculates the sum of the per-second rate of the HTTP requests over the last 5 minutes.
In this example, $http\_requests\_total[5m]$ is a range vector that represents the HTTP requests number time series over the last 5 minutes.
$rate$ is a function used to calculate the per-second average rate of the time series, and $sum$ is an aggregation operator used to calculate the sum over dimensions.
Readers can refer to \cite{promql} for more details of the PromQL.

\subsection{Motivation}
In practice, engineers may write PromQL queries to query metric data for different purposes, such as observing resource consumption trends, locating root causes, etc.
These PromQL queries can be complex, and writing them requires a sufficient understanding of the syntax of PromQL and the context of the system (e.g., the deployment of service instances).
This makes writing PromQL queries a time-consuming task.
In this paper, we aim to design an approach that can transform engineers' natural language questions into PromQL queries.
This approach encourages engineers to interact with Prometheus metrics through natural language in a more convenient and fast manner.

Figure~\ref{fig:motivation example} shows an example of a natural language question and the corresponding PromQL query.
In this example, the engineer wants to know which node has the most available memory among the nodes where the order service is deployed, so that he can decide the subsequent deployment strategy.
The corresponding PromQL query uses $topk$ operator to analyze the available memory metric of the corresponding node to achieve this purpose.
Since LLMs exhibit promising natural language understanding and code generation capabilities, it is straightforward to implement the approach based on LLMs.
However, we found it still has the following challenges:

\begin{figure}[t]
	\centering
	\includegraphics[width=0.7\linewidth ]{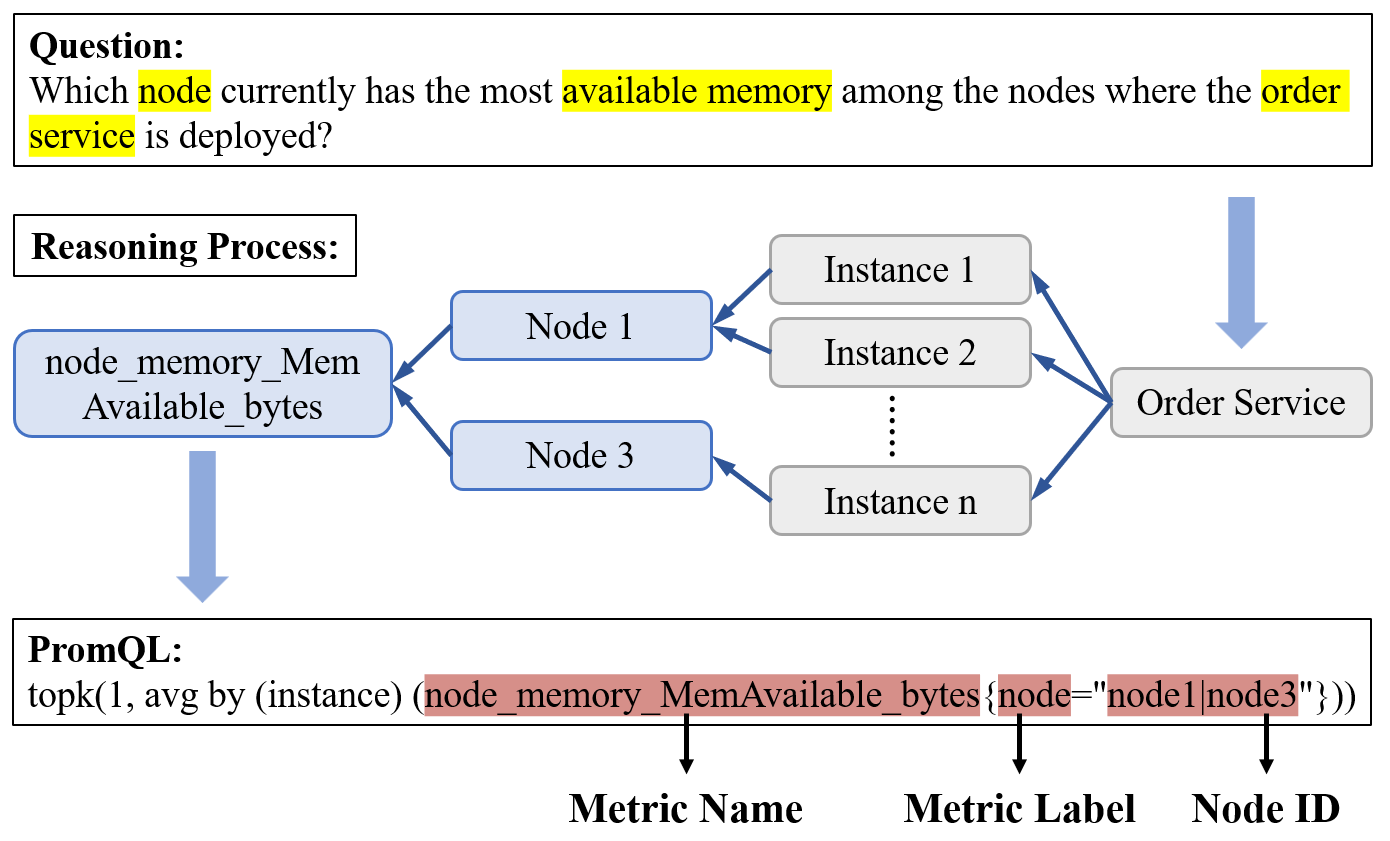}
	\caption{A Motivation Example}
	\label{fig:motivation example}
\end{figure}

First, generic LLMs are typically trained on public corpus, and when used in private domains, they inevitably encounter the issue of lacking private domain knowledge (e.g., metrics and component information in a real system).
This makes it difficult to generate PromQL queries that fit the system context by directly using public LLMs.
As shown in the example in Figure~\ref{fig:motivation example}, in order to generate this PromQL query, LLMs first need to know the syntax of PromQL and the operators and functions provided by PromQL.
LLMs have learned these knowledge through the public corpus during model training.
However, as shown in the red segment of the PromQL query in Figure~\ref{fig:motivation example}, generating this PromQL query also requires knowing the metrics name, metrics label, and Node ID, which is usually different in different systems. 
Both the metric name and label name can be customized by the engineer, and the node ID is represented differently in different systems.
Therefore, directly using LLMs to generate a PromQL query is likely to get the PromQL query which is syntactically correct, but with the wrong context information.

Second, approaches based on fine-tuning or few-shot learning from historical Q\&A samples are not suitable for text-to-PromQL.
Fine-tuning and few-shot learning have been widely used in a variety of AIOps tasks (e.g., log parsing~\cite{divlog,lilac}, root cause identification~\cite{few-shot-learning-rca,llm-rca-icse23,RCACopilot}, query recommendation~\cite{xpert}, etc.). 
These approaches use historical data to fine-tune LLMs or retrieve similar historical cases to enhance the quality of generated results.
In our task, generating PromQL queries often require knowledge of system components (e.g., containers, pods, services).
However, in modern online service systems (typically based on microservices architecture and deployed on the cloud), these system components are often dynamically changing.
As the example shown in Figure~\ref{fig:motivation example}, the nodes where the order service is deployed may change dynamically due to service scaling.
Therefore, the knowledge of system components in historical data may be difficult to match with the current system state.
Moreover, in real systems, only historical PromQL query statements can be easily collected, constructing high-quality Q\&A pairs still requires a lot of work from engineers, which is expensive and difficult.

Third, generating PromQL queries often involves a multi-hop reasoning process, which cannot be met by a general RAG (Retrieval-Augmented Generation) strategy.
General RAG strategies employ semantic retrieval or keyword retrieval methods to retrieve task-relevant knowledge from databases and incorporate it into the input prompt, thereby augmenting the output generated by LLMs.
In our task, due to the complexity of modern online service systems, it is difficult for engineers to know the real-time status of all the components of the system.
Therefore, the questions posed by engineers often do not contain all the information needed for writing PromQL queries.
As the example shown in Figure~\ref{fig:motivation example}, the reasoning process for writing this PromQL query is an obvious multi-hop reasoning process.
In the question, the engineer only provides the name of the service, but what it wants to know is the metrics of all the nodes that have deployed the service.
In order to find those nodes, it needs to first find all the service instances of that service and on which nodes those service instances are deployed.
It can be seen that it is difficult to obtain these multi-hop knowledge with a simple retrieval or matching method based on the information contained in the question.

Through the above analysis, we can find that in order to generate the PromQL query based on the natural language question, we need to provide knowledge of the metrics and system components related to the question to LLMs on demand.
At the same time, we need to be able to support multi-hop reasoning and retrieval to overcome the problem of insufficient original information.
Therefore, we use a knowledge graph to describe the complex context of a system.
Then, we achieve text-to-PromQL through the synergistic reasoning of knowledge graph and LLMs.








\section{Approach}
\label{section:framework}

\subsection{Overview}
In this work, we present a text-to-PromQL framework \app, which takes a natural language question as input and outputs the corresponding PromQL query of the question through synergistic reasoning with LLMs and knowledge graph.
The main idea of \app is to describe a specific system's context (e.g., metrics information, component dependencies) through a knowledge graph, and then adopt a knowledge graph-enhanced RAG paradigm for accurate PromQL query generation.

\begin{figure}[t]
	\centering
	\includegraphics[width=\linewidth ]{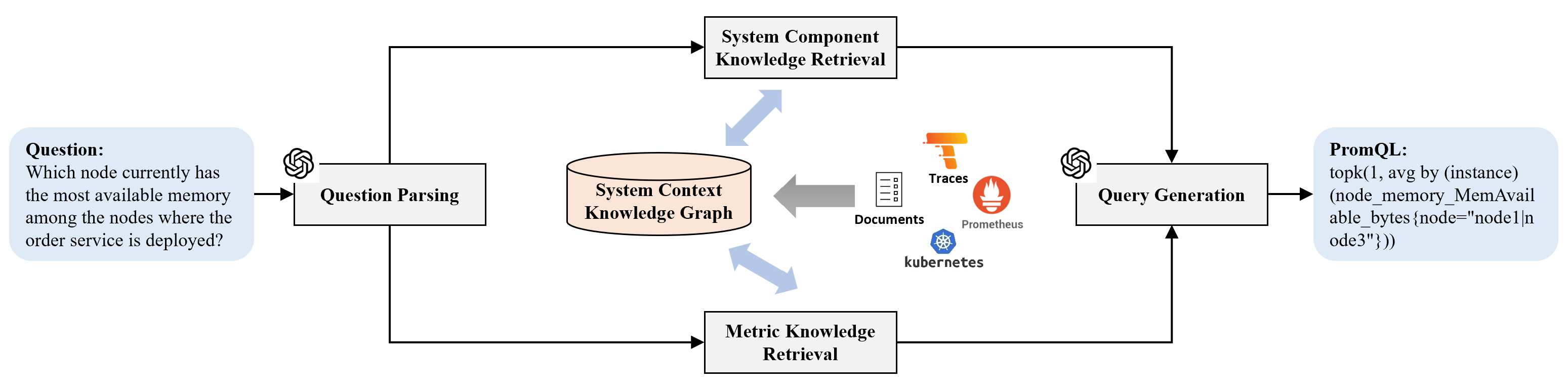}
	\caption{Overview of \app}
	\label{fig:overview}
\end{figure}

An overview of \app is shown in Figure~\ref{fig:overview}, the whole process includes two phases, i.e., system context knowledge graph construction and PromQL query generation from natural language questions.
In order to provide the system context needed for LLMs to generate PromQL queries, we first construct a system context knowledge graph based on system documents, traces, and metadata from Kubernetes and Prometheus.
Based on the system context knowledge graph, we convert natural language questions into PromQL queries through four steps, i.e., question parsing, system component knowledge retrieval, metric knowledge retrieval, and query generation.
Given a natural language question as input, \app first extracts the metrics and system components information from the question.
Then, based on the extracted metrics and component information, \app retrieves the system component knowledge and metric knowledge required to answer the question from the system context knowledge graph.
Finally, \app feeds both the initial natural language question and the retrieved knowledge into the LLMs to obtain the corresponding PromQL query.
To help readers understand the whole process of \app, we will use the example in Figure~\ref{fig:motivation example} in subsequent sections to illustrate how each step of \app works.

\subsection{System Context Knowledge Graph}
In this section, we introduce the definition of the system context knowledge graph and how we automatically construct the knowledge graph.
The knowledge graph proposed in this paper is designed for the current commonly used kubernetes~\cite{kubernetes}-based cloud native system architecture.

\begin{figure}[t]
	\centering
	\includegraphics[width=0.6\linewidth ]{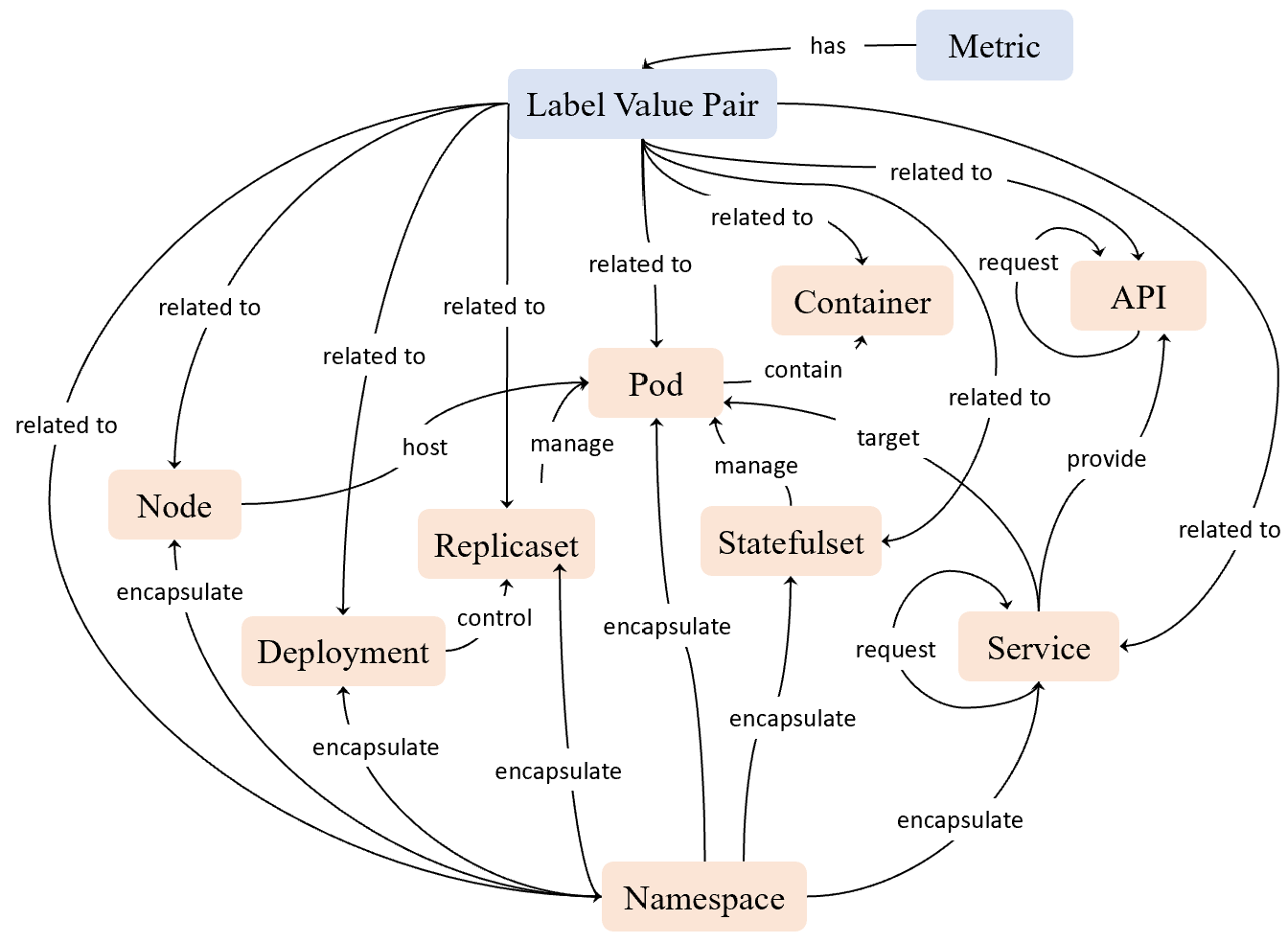}
	\caption{Schema of the System Context Knowledge Graph}
	\label{fig:schema}
\end{figure}

\subsubsection{Schema of the System Context Knowledge Graph}
Online service systems usually contain a variety of system components such as nodes, services, APIs, etc, and there are different relationships between these components.
Each monitoring metric may also be associated with one or more system components through different labels.
We want to describe the above relationships through the system context knowledge graph.
Therefore, we design the schema of the knowledge graph as shown in Figure~\ref{fig:schema}, where rounded rectangles indicate different types of entities in the knowledge graph and arrows represent the relationships between different types of entities.
In particular, the schema contains two kinds of entity types, system component entities (i.e., orange rectangles) and metric data entities (i.e., blue rectangles).

The system component entities include nine entity types, i.e., \textit{Node, Deployment, Namespace, Relicaset, Pod, Statefulset, Service, Container, API}.
These entity types are derived from the resources defined by Kubernetes~\cite{kubernetes} and the common concepts in microservice systems.
Relationships between different entity types have different practical meanings, for example: dependencies between resources, such as a pod containing multiple containers; hierarchical relationships between concepts, such as a service providing multiple APIs; and invocations relationships, such as one service requesting another.

The metric data entities include two entity types, i.e., \textit{Label-Value Pair, Metric}.
These two entity types are derived from the data model in Prometheus~\cite{prometheus}.
The metric entity is only connected to the label-value pair entity, whereas the label-value pair entity may be linked to specific system components.
With these two types of entities, we can identify one or more time series that we want to analyze.

\subsubsection{System Context Knowledge Graph Construction}
\label{kg_construct}
To construct the knowledge graph, we automatically extract various entities and relationships from different sources and fuse them together.
Specifically, we used four data sources to build the knowledge graph, and because these data are structured, we only need to design rules to enable their fusion and automate the whole process.
The details of how each data source is used are as follows:

\begin{itemize}
    \item \textbf{Prometheus}: Metadata of the monitored metrics in the system is stored in the server of Prometheus, which serves as the source of metrics knowledge. Specifically, we obtain the metric name, metric description, metric type, label name, and label value through the API provided by Prometheus. 
    These information are then used to construct the label-value pair entities and metric entities and their relations. 
    Specifically, we treat each metric as a metric entity in the knowledge graph, with its name serving as the entity's identifier and its type and description as the entity's attributes. Each unique label-value pair is treated as a label-value pair entity and associated with its corresponding metric entity.
    \item \textbf{Kubernetes}: Kubernetes stores the metadata about the resources it manages and provides APIs for querying the metadata. And the data model of Kubernetes already defines clear relations for the resources it manages. Therefore, we directly use these APIs get all the resources and their relations in the system.
    These resources correspond to the system component entities in the knowledge graph. We treat each resource as an entity in the knowledge graph, using its name as the entity's identifier and determining its entity type based on its resource type. We then construct edges between entities according to the relationship types defined in the schema and the actual relationships among the resources.
    \item \textbf{Traces}: A trace is the description of the detailed execution process of a request through the service instances in the system \cite{opentelemetry}. Through traces, we extract the services entities and APIs entities in the system and the request relations between services or APIs. Then, based on the resource-related attributes recorded in the span, we identify the relationships between service entities and other system components, and link them in the knowledge graph.
    \item \textbf{Documents}: In online service systems, engineers usually write documents to describe the functions of the services and APIs. These descriptions can help us recognize the intent of users' questions. Therefore, we extract the service descriptions and API descriptions from the system's documents and use them as the attributes of the corresponding entities.
\end{itemize}

After all the knowledge (including entities and their relationships) has been extracted from different sources, we link the label value pair entities to the corresponding system component entities based on the label name, label value and entity name.
After performing the above operations, we can obtain the system context knowledge graph.
Also, we can update the knowledge graph through periodic queries or proactive notifications, making it consistent with the system status all the time.
For the convenience of subsequent retrieval, we store the constructed knowledge graph in a graph database, and the names and descriptions of metrics, services, and APIs in a NoSQL database.


\subsection{Question Parsing}

As shown in the example in Figure~\ref{fig:motivation example}, in order to write the corresponding PromQL query, we need to know the detailed information of the metric and the system components in the question (e.g., metric name, metric label, and node ID, etc.).
We can see that all the required information is contained in the constructed system context knowledge graph.
However the graph is too large to feed the entire graph into LLMs, thus we only retrieve knowledge related to the question in the graph.
Specifically, we first parse the original question to obtain the raw information about the metrics and system components that the engineer queried.
Then we use them as the input to retrieve the relevant knowledge from the knowledge graph.
Note that because of the different characteristics of the metrics and system components, we parse and retrieve them separately.

Specifically, \app uses LLMs for question parsing since LLMs exhibit a powerful natural language understanding capability.
For each question \app prompts LLMs to extract components relation paths and metrics from the original question separately.
We then explain the extraction process in detail.

\begin{figure}[t]
	\centering
	\includegraphics[width=0.85\linewidth ]{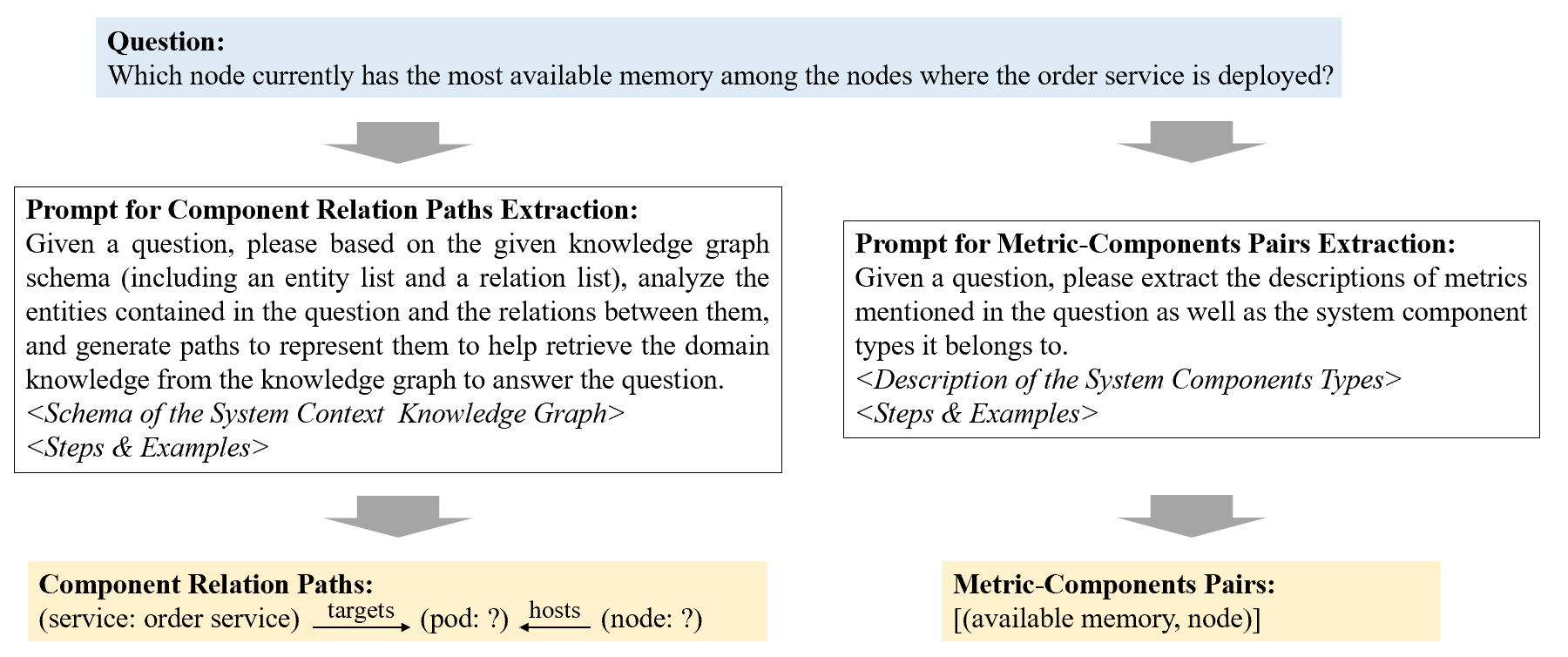}
	\caption{Example of the Prompt and Result of Question Parsing}
	\label{fig:question parsing}
\end{figure}

\subsubsection{Components Relation Paths Extraction}
In the questions from the engineers, they usually do not directly mention the components used in writing the PromQL query, but instead describe the dependencies between the target component and other components. 
As shown in the example in Figure~\ref{fig:motivation example}, instead of directly providing the ID of the target node, the engineer mentions that he is concerned with the node where the order service is deployed.
Therefore, for more precise knowledge retrieval, we extract paths from the question to represent the entities and their relationships involved in the question, which we call components relation paths.
Each components relation path is a sequence of entities and relationships $W=[e_1, r_1, e_2, r_2, \dots, r_{n-1}, e_n]$, where $e_i$ denotes the $i$-th entity and $r_i$ denotes the $i$-th relation in the relation path.
For each entity $e_i$, we keep its entity type and name; when the entity name is not extracted, we use a placeholder (e.g., $?$) to represent its name.

We design the prompt based on the Chain-of-Thought (CoT)~\cite{cot} and few-shot learning~\cite{few-shot-learning} paradigm.
We have designed a chain of thought for path extraction, as well as several fixed examples (5 in this paper), and included them in the prompt.
Meanwhile, to enable the LLMs to recognize the predefined entities and relations, we incorporate the schema of the system context knowledge graph in the prompt.
Figure~\ref{fig:question parsing} shows the designed prompt and the example component relationship path extracted from the question in Figure~\ref{fig:motivation example}.
It can be seen that the engineer in the original question is concerned with the node where the order service is deployed, and the path we extracted is $(service: order \; service) \xrightarrow{targets} (pod:?) \xleftarrow{host} (node:?)$.
This is because the service is deployed to a node as the pod, thus the path we extracted contains two hops.
Moreover, since the names of the pod and node are not mentioned in the question, they are represented as placeholders in the path.
Note that we instruct the model not to output paths where the metric names are all empty, and for some complex questions, we may extract multiple components relation paths.

\subsubsection{Metric-Component Pairs Extraction}
In the questions from the engineers, they usually describe the metrics they want to query and the system components associated with those metrics.
As shown in the example in Figure~\ref{fig:motivation example}, the engineer wants to query the available memory of the node.
Therefore, we extract the description of the metrics and the directly associated system components from the question.
We represent the parsing result as a set of pairs of metric and component type $M=[(m_1,c_1),(m_2,c_2), \dots,(m_n,c_n)]$, where $m_i$ denotes the discription of the $i$-th metric and $c_i$ denotes the component type directly associated with the $i$-th metric.

We also design the prompt based on the CoT and few-shot learning paradigm.
Figure~\ref{fig:question parsing} shows the prompt template and the example metric-component pair extracted from the question in Figure~\ref{fig:motivation example}.
The pairs of metric and component type extracted from this question is $[(available\; memory, node)]$.
This result indicates that the engineer wants to query the available memory at the node level.
Note that if the model does not extract a metric description, we will use the whole sentence of the question as the metric description. 
If the component type that the metric belongs to is not extracted, we will set the component type as "ALL", which corresponds to all components in the system.

\subsection{System Component Knowledge Retrieval}
In this section, we introduce how \app uses the extracted components relation paths to retrieve question-related system component knowledge in the knowledge graph.
Our main idea is to retrieve path instances in the knowledge graph that meet the requirements (including entities, and relations) of components relational paths, and we call them reasoning paths.
Existing KG-based LLM reasoning approaches~\cite{ToG, RoG} only extract entities or paths consisting of entities from the question as the input of the retrieval step, while the component relationship paths in our work contain both entities and relationships.
Therefore, we extend the path retrieval methods in existing approaches~\cite{ToG, RoG} to consider both entities and the relationships between entities during retrieval.
Specifically, given a set of components relation paths, \app uses the following steps to retrieve the reasoning paths.

First, we preprocess the components relation paths to ensure that each path starts with a specific entity.
For the path with the first entity name is a placeholder, we look for the first entity in the path whose name is not a placeholder, and use this entity as the starting point to split the path into two paths.
Moreover, since the entity names extracted from the original question may be different from the real entity names, we replace them with the real entity names by similarity search.
Specifically, for each entity in the path with an entity name, we use BM25 algorithm \cite{bm25} to retrieve the names of all entities of the same type as that entity in the knowledge graph, and then use the most similar entity name to replace it.
For example, the original service name extracted from the question is $ticket \;  management \;  service$, which will be replaced with the actual service name $ticket \;  service$ after the above steps.

Then, for each preprocessed components relation path, we adopt a breadth-first search algorithm to retrieve the reasoning paths in the knowledge graph.
The pseudocode of the algorithm is presented in Algorithm~\ref{bfs}.
The algorithm first initializes the reasoning path with the first entity of the components relation path.
Then it iteratively extends each reasoning path by searching for eligible relations and entities.
For entities whose entity name is a placeholder, all entities with the same entity type as the entity are used to extend the reasoning path.
After all components relation paths are processed, we get a set of reasoning paths.
For the example shown in Figure~\ref{fig:motivation example}, we can get $n$ (number of service instances) reasoning paths, such as $(service: order \; service) \xrightarrow{targets} (pod:pod1) \xleftarrow{host} (node:node1)$ and $(service: order \; service) \xrightarrow{targets} (pod:pod3) \xleftarrow{host} (node:node3)$, where pods correspond to different service instances.
It can be seen that these reasoning paths demonstrate the real deployment relationships of $order \; service$.

\begin{algorithm}[t]
\caption{Retrieve system component knowledge based on components relation path}\label{bfs}
\begin{algorithmic}[1]
\Require  Knowledge Graph $G$, Components Relation Path $P= [e_1, r_1, e_2, r_2, \dots, e_n]$
\Ensure  Reasoning Paths $Paths$




\State $Paths \gets [[e_1]]$ \Comment{Initialize paths with the start node}
\For{$index \gets 1$ \textbf{to} $\text{length}(P) - 1$ \textbf{step} 2}
    \State $relation \gets P[index]$
    \State $expectedEntity \gets P[index + 1]$
    \State $newPaths \gets []$
    \For{$path$ \textbf{in} $Paths$}
        \State $currentNode \gets path[-1]$
        \State $connectedNodes$ $\gets$ $GetNodesByRelation($
        \Statex \quad\quad\quad $currentNode, relation)$
        \For{$node$ \textbf{in} $connectedNodes$}
            \If{$node = expectedEntity \quad \textbf{or} $ 
            \Statex \quad\quad\quad\quad   $expectedEntity.name = '?'$}
            \State $newPath \gets path + [relation, node]$
            \State $newPaths.append(newPath)$
            \EndIf
        \EndFor
    \EndFor
    \State $Paths \gets newPaths$
\EndFor

\State \Return $Paths$
\end{algorithmic}
\end{algorithm}

\subsection{Metric Knowledge Retrieval}
In this section, we introduce how \app uses the extracted metric-component pairs to retrieve the metrics and label-value pairs related to the question.
Our main idea is to determine the range of candidate metrics by the component type of each metric in the metric-component pairs and select the most relevant metrics from them.
Then we recognize the relevant label-value pairs based on the reasoning path retrieved in the previous step.

\subsubsection{Metrics Retrieval}
For each metric-component pair, we first retrieve all entities whose entity type is the same as the component type in the metric-component pair. 
Then, we search in the knowledge graph for all metrics entities that are reachable after two hops from these entities, i.e., all metrics which are associated through label-value pair entities.
Subsequently, for these metrics, we use the BM25 algorithm~\cite{bm25} to compute and rank the correlation between the metric description in the metric-component pair and the descriptions of these metrics themselves.
And we treat the top $k$ metrics ($k=10$ in this paper) among them as the candidate metrics.
We select multiple metrics because the metrics queried by users may be derived from the calculations of several real metrics (such metrics are typically referred to as derived metrics). 
For example, the request error rate is a derived metric that is calculated based on the total number of requests and the number of request errors.
Based on our observations, a single derived metric usually involves 2-5 real metrics.
We recommend that users determine the value of $k$ based on the system's scale and the observations  of the derived metrics.
Suppose that $n$ metric-component pairs are obtained after question parsing, then we will obtain no more than $nk$ candidate metrics in total.
For some complex problems, we may get a large number of candidate metrics, which may make the final prompt exceed the token limit.
Therefore, we prompt LLMs to select several most relevant metrics from the candidate metrics as the final retrieval result.
And we adopt the same prompt design approach as in the question parsing phase.
For the example shown in Figure~\ref{fig:motivation example}, we can get the metric $node\_memory\_MemAvailable\_bytes$ after the above steps.

\subsubsection{Label-Value Pairs Retrieval}
For each metric we retrieved, we further retrieval the label-value pairs that are relevant to the question.
In real systems, the metric labels can be divided into two categories, the first is related to the system components, such as pod, service, etc., and the second is related to the semantic of the metric, such as method, mode, etc.
We retrieve the two types of metric labels separately.

For the labels related to the system components, the reasoning paths we retrieved in the previous step already contain the components related to the question.
Therefore, for each retrieved metric, we directly retrieve the paths that are starting with the metric and ends with the components in the reasoning paths, which are in the form of $metric \xrightarrow{} label{-}value \; pair \xrightarrow{} component$.
We treat these paths as the knowledge of the component related labels and add them to the reasoning paths.

For the labels related to the semantic of the metric, we use LLMs to select the required label-value pairs.
For each metric, we first obtain the labels related to the semantic of the metric from the knowledge graph, i.e., the labels corresponding to the label-value pair entities which are not connected to any component entities.
Then, we input the name, description, and type of each metric, and the name and example values of the labels into the LLMs.
And we prompt the LLMs to identify the labels that are related to the question and to generate descriptions of the label values.
Subsequently, based on the obtained descriptions of each label value, we use BM25 algorithm~\cite{bm25} to retrieve the relevant label-value pair entities and take the top $m$ entities as the final result.
We then represent the result as paths in the form of $metric\xrightarrow{}label{-}value \; pair$ and add them to the reasoning paths.

For the example in Figure~\ref{fig:motivation example}, we can get two paths, $(metric:$ $node\_memory$ $\_MemAvailable\_bytes)$ $\xrightarrow{has}$ $(metric\_label$ $\_value:node=node1)$ $\xrightarrow{related \; to}$ $(node:node1)$ and $(metric:$ $node\_memory$ $\_MemAvailable\_bytes)$ $\xrightarrow{has}$ $(metric\_label$ $\_value:$ $node=node3)$ $\xrightarrow{related \; to}$ $(node:node3)$.
These two paths will serve as reasoning paths together with the retrieved system component knowledge paths.

\subsection{Query Generation}

\begin{figure}[t]
	\centering
	\includegraphics[width=0.55\linewidth ]{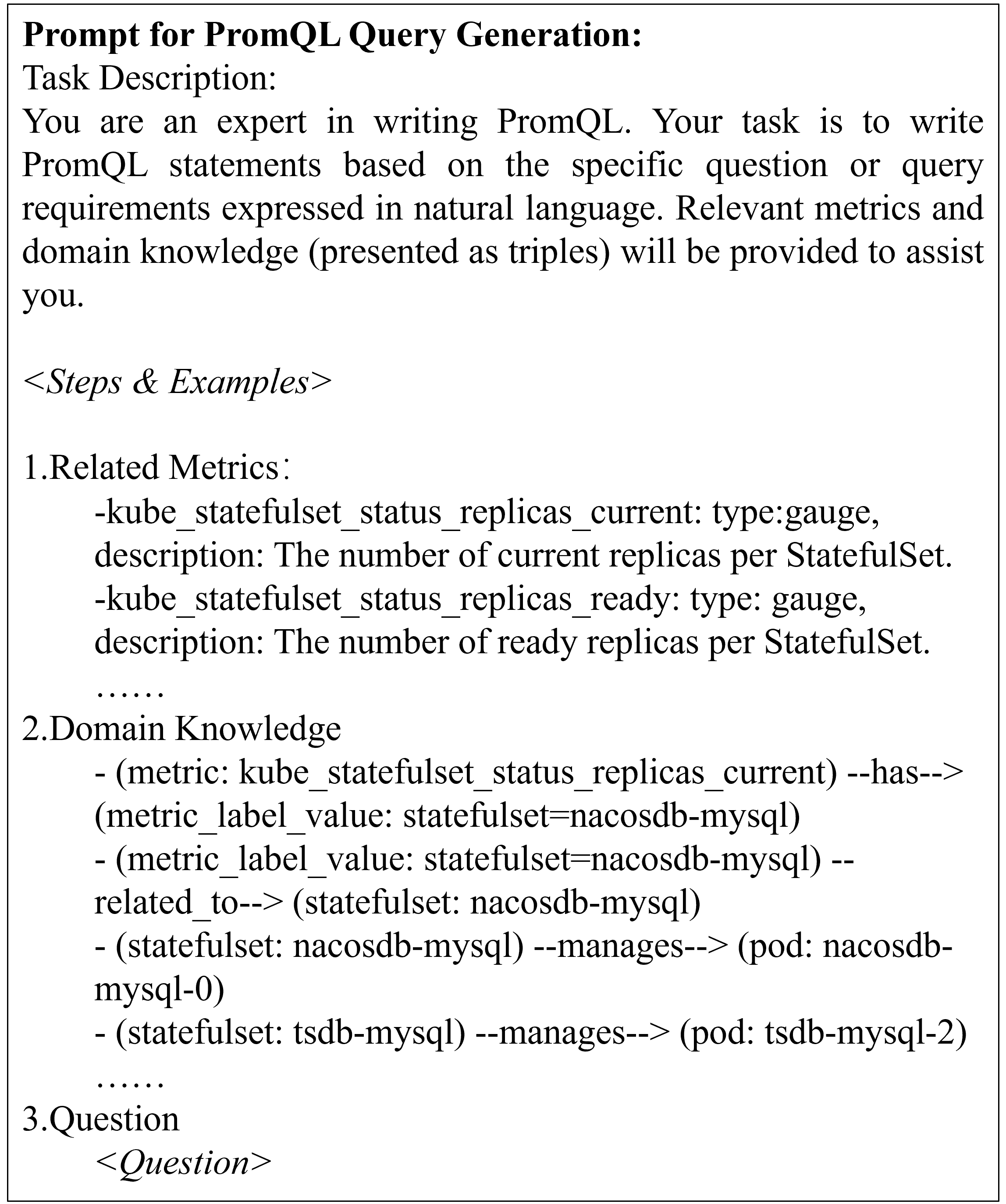}
	\caption{An Example of the Prompt for PromQL Query Generation}
	\label{fig:final prompt}
\end{figure}

Based on the retrieved system component knowledge and metric knowledge, \app leverages LLMs to transform the natural language question into a PromQL query.
An example of the prompt used for PromQL query generation is shown in Figura~\ref{fig:final prompt}, which is also designed based on the CoT and few-shot learning paradigm.
We included a designed cot and one fixed example in the prompt.
Additionally, as shown in the example we incorporate the retrieved knowledge into the prompt as two parts.
In the first part, there is the name, type, and description of the metrics we retrieved.
In the second part, there are all the reasoning paths we retrieved.
Note that since some of the knowledge is repeated in multiple reasoning paths, we split the reasoning paths into triples for representation.
The constructed prompt is directly fed to LLMs to get the corresponding PromQL query.

Due to space limitations, the details of all the prompts used by \app are included in the replication package~\cite{PromCopilot}.

\section{Evaluation}
\label{section:evaluation}
To evaluate \app we conduct a series of experiment studies to answer the following research questions:

\begin{itemize}
    \item \textbf{RQ1:} How effective is \app in translating natural language question to metric query?
    
    \item \textbf{RQ2:} How accurate is \app in system component knowledge retrieval and metric knowledge retrieval?

    \item \textbf{RQ3:} How much does the different knowledge contribute to the effectiveness of \app?

    \item \textbf{RQ4:} How useful is \app in helping engineers write PromQL queries?
\end{itemize}

\subsection{Dataset}
To the best of our knowledge, there are no public datasets for the text-to-PromQL task.
To fill this gap, we manually construct the first text-to-PromQL benchmark dataset based on an open-source microservice-based online service system.
Our benchmark dataset contains 280 manually constructed question-PromQL query pairs.
The most complex PromQL query in the dataset contains 47 tokens (including metric name, metric label, metric label value, operator, function, etc.).
Meanwhile, we run the system for a week and include in the dataset the metrics data, trace data, and system deployment records generated during this period.
Researchers can use these data to quickly restore the state of the system and perform metric queries, and we believe it is helpful for subsequent research.
The benchmark dataset is publicly available at~\cite{PromCopilot}.

\subsubsection{Metric Collection}
We use the latest release v1.0.0 of the medium-scale open-source microservice system TrainTicket~\cite{trainticket-tse,trainticket_poster} to construct our benchmark dataset.
TrainTicket is an online train ticket booking system and has been widely used in researches on microservice architecture~\cite{microeval}, anomaly detection~\cite{deeptralog}, and root cause analysis~\cite{traceContrast, eadro, nezha}.
It contains more than 40 services implemented in different languages.
We deploy the system to simulate a real metrics collection and query environment.

We first deploy TrainTicket on a Kubernetes cluster and use Prometheus to collect metrics data.
The cluster contains six nodes, each of which has a 16-core 3.0GHZ CPU, 32GM RAM, and runs with Ubuntu 20.10.
TrainTicket is deployed on these nodes for a total of 96 service instances.
We use four exporters to collect metrics from different components of the system, including node-exporter\cite{node_exporter}, cadvisor\cite{cadvisor}, kube-state-metrics\cite{kube_state_metrics}, and metrics-generator\cite{metrics_generator}.
These exporters predefine various metrics and specify the name, type, label, and description of each metric.
We also customize several metrics in addition to these predefined metrics.
All the metrics are collected and stored by Prometheus.
OpenTelemetry\cite{opentelemetry} and Grafana Tempo~\cite{tempo} are used as the distributed tracing framework to collect and store traces.

To produce metric data, we use the load generator provided by TrainTicket~\cite{trainticket-tse} to execute automated test cases to simulate user requests.
The whole process lasts for 7 days.
Then we collect all the metrics, traces, and Kubernetes deployment records generated during this period as the raw data of our benchmark dataset.
In total, we collect 209 metrics, and each metric contains at least 4 labels and at most 16 labels.
Table~\ref{tab:stat} shows the number of system components and metrics included in the dataset.

\begin{table}
	\caption{Number of System Components and Metrics in the Dataset}	\label{tab:stat}
  \centering
    \resizebox{0.5\linewidth}{!}{
		\begin{tabular}{c|c|c|c}
			\hline
			Name & Number & Name & Number   \\
			\hline
			API & 140 & Metric & 209 \\
			Pod & 147 & Label-Value Pairs& 2,489 \\
			Container & 163 & Node & 6 \\
			Deployment & 140 & Namespace & 8 \\
			Service & 77 & Replicaset & 55 \\
            Statefulset & 7 & & \\
			\hline
	\end{tabular}
 }
\end{table}

\subsubsection{Question-Answer Pair Construction}
In order to design realistic natural language query questions, we first investigate typical metric query usage scenarios.
Specifically, we collected typical PromQL query cases from multiple sources, including 33 cases from official documentation, 38 cases from third-party tutorials, and 45 cases from Stack Overflow (filtered from 500 related questions).
In addition, we also collect the top 200 distinct PromQL queries executed in a week from one of the teams at Alibaba Group.
These cases include simple and complex PromQL queries, and cover common scenarios such as average latency calculation, error rate calculation, metric sorting, metric prediction, etc.

Based on the collected typical cases, we ask two Ph.D students and a faculty member to analyse the cases and design the questions, all of whom had work or internship experience at world-leading ISPs.
We first summarize the collected cases into different typical scenarios, where PromQL queries with similar intent would be considered as the same scenario.
Then we summarize the natural language question for each scenario based on the description of the query in the collected documents or questions.
Finally, each participant then instantiates these scenarios into different natural language questions and corresponding PromQL queries in our system based on his/her understanding of the system and the question.
For each designed question, the other two participants will confirm its rationality. 
If a question is contentious, all participants will further discuss it until they reach a consensus.
Moreover, we invite five engineers (including two authors) from two teams in Alibaba Group to help us further confirm the soundness of the constructed questions, all of whom have more than five years of work experience.
Each engineer is assigned 80 questions, with some questions assigned to two engineers.
We then require each engineer to carefully review the questions and answers and execute queries on our Prometheus platform to verify the validity and correctness of the questions and answers.
For questions that the engineers consider questionable, we discuss the reasons for their concerns and make revisions accordingly.
Finally, we have designed a total of 280 questions, and crafted the corresponding PromQL query for each question.
In each question, the corresponding PromQL query may involve 1 to 4 metrics, 1 to 8 labels, and 1 to 10 operators or functions.
The most complex PromQL query in the dataset contains 47 tokens (including metric name, metric label, metric label value, operator, function, etc.).
These questions cover a portion of the entities in the system, including 98 metrics (46.89\%), 43 services (55.84\%), 87 pods (59.18\%), and others. 
For more details, please refer to the replication package~\cite{PromCopilot}.
Note that each question may have one or more corresponding PromQL queries, this is because PromQL queries using different operators or functions may achieve the same result.


\subsection{Evaluation Setup}

\subsubsection{Studied LLMs}


We select two LLMs in our experiments, i.e., \textbf{GPT-4-Turbo}~\cite{gpt4}, and \textbf{DeepSeek-Coder-V2-236B}~\cite{deepseek-coder-v2}, for the following reasons.
First, these LLMs have been widely used in previous software engineering research (e.g., code generation~\cite{codereval,icse24-code}, AIOps~\cite{xpert, RCACopilot,divlog}) and have shown advanced effectiveness.
Second, these LLMs contain both generic LLM (GPT-4-Turbo) and code LLM (DeepSeek-Coder-V2-236B), which can provide a more comprehensive view of the effectiveness of \app and baseline approaches.
We implement \app and all baseline approaches with each of the above LLMs.

\subsubsection{Baselines}
To the best of our knowledge, there is no research specifically on text-to-PromQL. 
Therefore we follow the typical practice and use the following two approaches as baselines.

\begin{itemize}
    \item \textbf{Basic Prompt}: This approach directly queries LLMs for text-to-PromQL, with a prompt designed based on the Chain-of-Thought (CoT)~\cite{cot} paradigm. The prompt is similar to the example prompt in Figure~\ref{fig:final prompt}, but excludes the domain knowledge and examples. It indicates the basic capabilities of LLMs in text-to-PromQL.
    \item \textbf{Few-shot Learning}~\cite{xpert}: This approach enhances the basic prompt with retrieval-based few-shot learning. We retrieve questions similar to the current question from the historical question-answer pairs, and include the similar question-answer pairs in the prompt. This approach is derived from Xpert~\cite{xpert}, which uses few-shot learning to recommend queries related to the incident in online service systems, and we modify it to match the text-to-PromQL task. We use \textbf{$\bm{k}$-shot} to represent this approach, where $\bm{k}$ represents the number of retrieved question-answer pairs (we set $\bm{k}$ as 1, 3 and 10 in the evaluation). 
\end{itemize}

It should be noted that all the prompts used by \app are equipped with fixed examples, and don't use historical question-answer pairs.
It makes \app significantly different from the few-shot learning-based baseline approach.
Due to page limitations, all the prompts and codes of the baseline approaches are included in our replication package~\cite{PromCopilot}.

\subsubsection{Evaluation Metrics}
We adopt the following three metrics to evaluate the effectiveness of \app and baseline approaches.

\begin{itemize}
    \item \textbf{MetricAcc}: The percentage of PromQL queries using the correct metrics out of all generated PromQL queries. We treat a PromQL query as using the correct metrics only if all the metrics used in that query are the same as the ground truth.
    For example, if a ground truth metric query uses metrics $a$ and $b$, then we consider the generated query using metrics correctly only if it uses exactly metrics $a$ and $b$.
    \item \textbf{SyntaxAcc}: The percentage of syntactically correct PromQL queries out of all generated PromQL queries. A PromQL query is considered syntactically correct only if it passes Promtheus' syntax checking.
    \item \textbf{QueryAcc}: The percentage of correct PromQL queries out of all generated PromQL queries. A PromQL query is considered correct only if the query is the same as the ground truth.
    Note that since a PromQL query may have multiple equivalent implementations, we consider all equivalent implementations as correct. 
    

\end{itemize}

It should be noted that $QueryAcc\leq MetricAcc$ and $QueryAcc \leq SyntaxAcc$, because a correct PromQL query requires both the syntax of the query and the metrics in the query are correct.
Furthermore, to ensure the correctness of the results, we manually checked the evaluation results of all metrics.

\subsubsection{Implementation}
We implement \app using Python 3.10, Neo4j 5.14.0~\cite{neo4j}, and ElasticSearch 8.11.0~\cite{es}.
We construct the system context knowledge graph based on the metric data, trace data, and deployment records in the dataset.
And we use the wiki of TrainTicket{~\cite{tt_wiki} as system documentation and extract API descriptions and service descriptions from it.
The final knowledge graph contains 3,356 entities and 43,405 relations, and we store it in the graph database Neo4j.
We store the names and descriptions of metrics, services, and APIs in ElasticSearch, which is a high-performance search engine that supports multiple search methods (such as BM25 and vector search) and has been widely used to build RAG systems~\cite{huggingface_es,OpenAI_es}.
We set $k$ in the metrics retrieval step as 10 based on the scale of the system (over 40 services and 209 metrics). 
We set $m$ in the label-value pairs retrieval step as 1.
We randomly select 50 questions from the dataset as historical questions to implement the few-shot learning approach. 
The remaining 230 questions are used as the test set for all approaches.
All the code and data can be found in our replication packege~\cite{PromCopilot}.

\begin{table}[t]
	\caption{Effectiveness of \app and Baseline Approaches}	\label{tab:rq1}
  \centering
  \resizebox{0.95\linewidth}{!}{
\begin{tabular}{c|c|c|c|c}
\hline
                                       LLM & Approach              & MetricAcc & SyntaxAcc & QueryAcc \\ \hline
\multirow{5}{*}{GPT-4-Turbo}                   & Basic Prompt          & 28.3\%      & 86.1\%       & 2.6\%      \\
                                        & Basic Prompt + 1-shot & 57.8\%      & 96.1\%      & 15.2\%     \\
                                        & Basic Prompt + 3-shot & 70.4\%      & 96.1\%      & 21.3\%     \\
                                        & Basic Prompt + 10-shot & 77\%      & 96.5\%      & 37.4\%     \\
                                        & \textbf{\app}   & 91.3\%      & 96.1\%      & 69.1\%     \\ \hline
\multirow{5}{*}{DeepSeek-Coder-V2-236B} & Basic Prompt          & 39.6\%      & 95.7\%      & 5.2\%     \\
                                        & Basic Prompt + 1-shot & 74.3\%      & 96.1\%      & 25.2\%     \\
                                        & Basic Prompt + 3-shot & 77.4\%     & 95.2\%      & 33\%     \\
                                        & Basic Prompt + 10-shot & 81.3\%      & 93\%      & 42.6\%     \\
                                        & \textbf{\app}   & 91.3\%      & 95.7\%      & 66.9\%     \\ \hline
\end{tabular}
}
\end{table}

\subsection{RQ1: Effectiveness}

\begin{figure}[t]
	\centering
	\includegraphics[width=0.7\linewidth ]{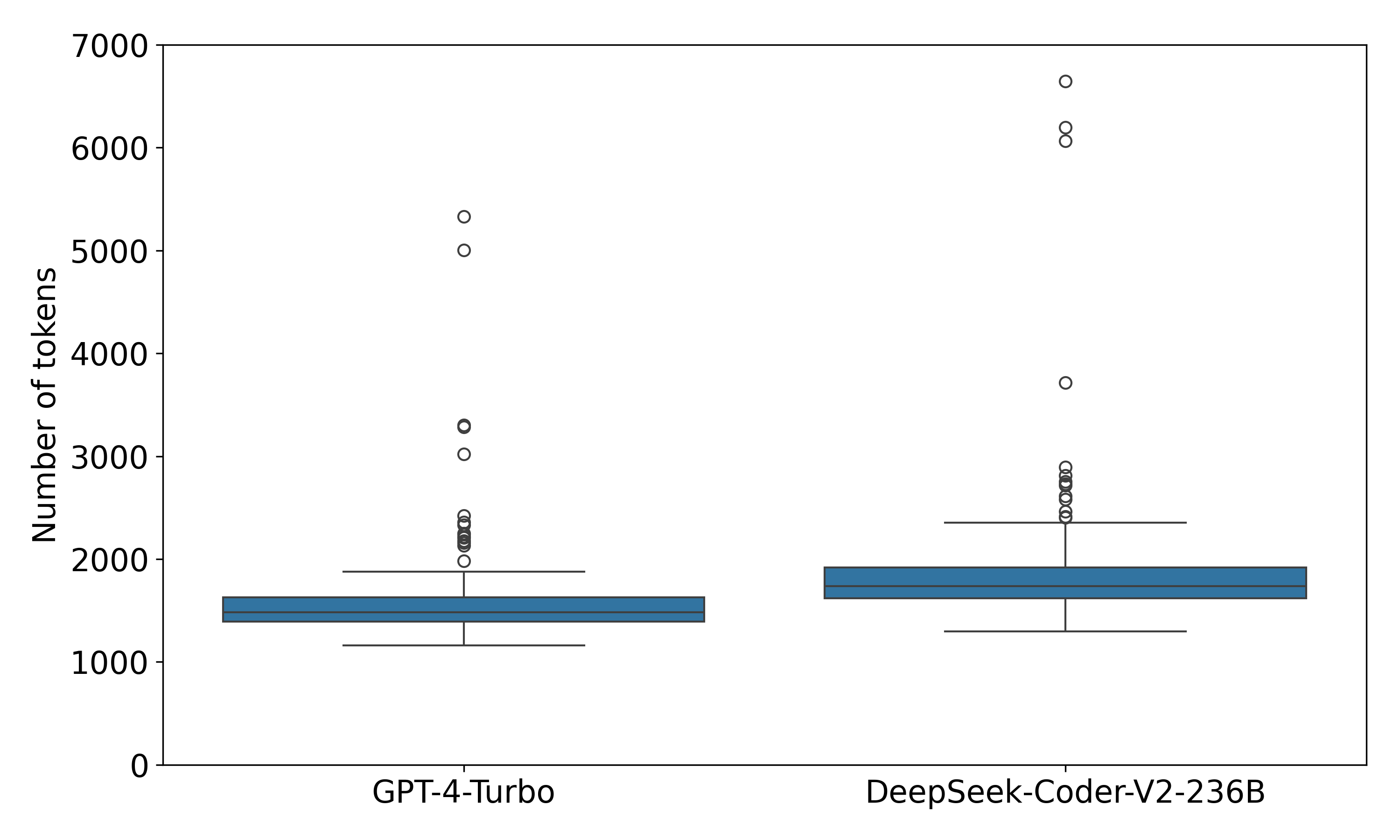}
	\caption{Token Consumption Using Different LLMs}
	\label{fig:token counts}
\end{figure}

Table~\ref{tab:rq1} shows the effectiveness evaluation results of \app and baseline approaches.
It can be seen that \app exhibits superior performance compared to all the baseline approaches, and achieves the highest MetricAcc (91.3\%), and QueryAcc (69.1\%).



It can be seen that although the code LLM DeepSeek-Coder-V2-236B has better code generation capability compared to GPT-4-Turbo, its performance is about the same as GPT-4-Turbo in our evaluation.
Specifically, MeticAcc and SyntaxAcc of all approaches are close.
And the QueryAcc for Basic Prompt and Few-shot Learning is slightly higher when using DeepSeek-Coder-V2-236B than when using GPT-4-Turbo.
This may be due to the better code generation capability of DeepSeek-Coder-V2-236B.
However, the QueryAcc of \app with DeepSeek-Coder-V2-236B (66.9\%) is lower than that with GPT-4-Turbo (69.1\%), which may be because the generic LLMs can better understand the contextual knowledge provided by \app.


The Basic Prompt shows the poorest performance among all approaches.
We find that while its SyntaxAcc reaches a maximum of 95.7\%, its QueryAcc only reaches a maximum of 5.2\%.
On the other hand, the MetricsAcc of the Basic Prompt reaches a maximum of 39.6\%.
This is because we use some open-source metric exporters when constructing the dataset, and the code and documents of these exporters may have been used for model training of the LLMs.
Thus, LLMs can correctly use a part of the metrics without additional knowledge.
The results of Basic Prompt demonstrate that the LLMs struggle to generate correct PromQL queries when lacking knowledge of metrics and system components.

As shown in the results, enhancing the Basic Prompt with few-shot learning improves the performance of generating PromQL queries. 
As the number of input historical examples increases, the QueryAcc also increases significantly, e.g., the QueryAcc increase from 5.2\% to 42.6\% when using DeepSeek-Coder-V2-236B.
This indicates that retrieving more historical examples can help improve the performance of PromQL query generation.
However, the performance of the approach based on few-shot learning is still lower than \app.
This is because historical examples cannot provide all the knowledge related to the question.
Especially for questions that have not appeared before, historical examples cannot help the LLMs to reason and generate the corresponding PromQL query.
Moreover, it is important to note that \app achieves better performance without using historical examples.
In practice, the performance of \app can be further improved through few-shot learning or fine-tuning.

Moreover, we find that the QueryAcc of all approaches shows a decrease compared to its MetricCcc and SyntaxAcc.
For example, when using GPT-4-Turbo \app achieves 91.3\% and 96.1\% in terms of MetricAcc and SyntaxAcc, but only 69.1\% in terms of QueryAcc.
This is probably due to two reasons. 
First, in addition to using the correct metrics, generating the correct PromQL query requires knowledge of the metric labels and system components related to the question.
Second, we found that even if all the required knowledge is provided to the LLMs, the genetated PromQL queries are not always correct.
This is because LLMs also need to understand and use the provided knowledge properly, and this process may also cause errors.

For LLM-based approaches, the token consumption is critical. If too many tokens are input it may result in the prompt exceeding the context length limit, thus obtaining incorrect results.
Therefore, as shown in Figure~\ref{fig:token counts}, we statistic the token consumption of \app when using different LLMs.
It can be seen that in most test cases, the number of consumed tokens is less than 2000. 
Only a few test cases consume a higher number of tokens, but the maximum is less than 7000.
DeepSeek-Coder-V2-236B consumes more tokens because its tokenizer produces more tokens for the same text than GPT-4-Turbo's tokenizer.
The maximum context length of both GPT-Turbo and DeepSeek-Coder-V2-236B is 128k, it can be seen that in the experiments no test case exceeds the limit.
Considering that our experiments are based on a medium-sized online service system (containing 147 pods and 77 services), it is reasonable to believe that \app can be used in larger scale systems in the industry.

In conclusion, \app is effective in translating natural language questions to PromQL queries and outperforms Basic Prompt and few-shot learning-based approaches.
And \app does not consume a lot of tokens, which demonstrates the cost-effectiveness of \app.
Compared to the baseline approaches, \app improved the QueryAcc from 37.4\% to 69.1\% and from 42.6\% to 66.9\% when using GPT-4-Turbo and DeepSeek-Coder-V2-236B, respectively.


\subsection{RQ2: Knowledge Retrieval Accuracy}

\begin{table}[t]
	\caption{Accuracy of Metrics Retrieval}	\label{tab:rq2-metric}
  \centering
  \resizebox{0.9\linewidth}{!}{
\begin{tabular}{c|c|c|c}
\hline
Approach & Avg Precision & Avg Recall &Avg F1-Score \\ \hline
\app (GPT-4-Turbo)                    & 0.907                         & 0.903  & 0.901    \\
\app (DeepSeek-Coder-V2-236B)                 & 0.857                         & 0.882  & 0.864    \\
\hline
\end{tabular}
}
\end{table}

\begin{table}[t]
	\caption{Accuracy of Reasoning Paths Retrieval}	\label{tab:rq2-triples}
  \centering
  \resizebox{0.9\linewidth}{!}{
\begin{tabular}{c|c|c|c}
\hline
Approach & Avg Precision & Avg Recall & Avg F1-Score \\ \hline
\app (GPT-4-Turbo)                    & 0.691                         & 0.908  & 0.741    \\ 
\app (DeepSeek-Coder-V2-236B)                 & 0.656                         & 0.864  & 0.695    \\
\hline
\end{tabular}
}
\end{table}


The accuracy of knowledge retrieval is crucial to the effectiveness of \app. 
Therefore, in this section we evaluate the accuracy of \app in retrieving different knowledge.
We first label the ground truth of the metrics and triples that are necessary in the query generation step for each question.
Then we calculated the precision, recall, and F1-score for each question, based on its actual retrieved metrics and triples.
Finally, we averaged the results from all questions to get the average precision, average recall, and average F1-score.

Table~\ref{tab:rq2-metric} presents the result of the accuracy of \app in metrics retrieval.
Note that we only statistics whether the correct metrics were retrieved (corresponding to the related metrics in Figure~\ref{fig:final prompt}), without considering the metric labels.
And retrieving the correct metrics does not mean that the metrics in the generated PromQL query are also correct.
As shown in Table~\ref{tab:rq2-metric}, all of the average F1-score exceed 0.86.
In particular, when using GPT-4-Turbo the average precision and average recall exceed 0.9.
This indicates that \app can effectively understand the intent of the question and retrieve the correct metrics.

Table~\ref{tab:rq2-triples} presents the result of the accuracy of \app in reasoning paths retrieval.
Note that this result is obtained by statistics on the correctness of the retrieved triples, corresponding to the domain knowledge in Figure~\ref{fig:final prompt}.
Although the average precision in the results is relatively low (the highest only reaches 0.691), the average recall both exceed 0.86.
This result is in line with the desired objective of our approach.
\app has retrieved some knowledge that is not related to the question, but the vast majority of the required knowledge has been retrieved by \app.
This retrieval result ensures that \app can generate the correct PromQL query.

In conclusion, \app is accurate in system component knowledge retrieval and metric knowledge retrieval.
It achieves a high recall in both metrics retrieval (0.903) and reasoning paths retrieval (0.908).

\begin{table}[t]
	\caption{Evaluation of Contribution of Different Knowledge}	\label{tab:rq3}
  \centering
  \resizebox{0.95\linewidth}{!}{
\begin{tabular}{c|c|c|c|c}
\hline
LLM                                    & Approach  & MetricAcc & SyntaxAcc & QueryAcc \\ \hline
\multirow{3}{*}{GPT-4-Turbo}                   & \app w/oMK & 38.7\%      & 90.9\%      & 27.8\%     \\
                                        & \app w/oSK & 90.9\%      & 95.2\%      & 12.2\%     \\
                                        & \textbf{\app}   & 91.3\%      & 96.1\%      & 69.1\%     \\ \hline
\multirow{3}{*}{DeepSeek-Coder-V2-236B} & \app w/oMK & 39.6\%      & 96.9\%      & 26.5\%     \\
                                        & \app w/oSK & 87.4\%      & 95.7\%      & 10.4\%     \\
                                        & \textbf{\app}   & 91.3\%      & 95.7\%      & 66.9\%     \\ \hline
\end{tabular}
}
\end{table}

\subsection{RQ3: Ablation Study}
We perform an ablation study to evaluate how different knowledge contributes to the effectiveness of \app.
We derive two variants of \app called \textbf{\app w/oMK} and \textbf{\app w/oSK}.
\app w/oMK removes the metric knowledge retrieval module from the approach, and only uses the system component knowledge to generate the PromQL query.
\app w/oSK removes the component knowledge retrieval module from the approach, and only uses the metric knowledge to generate the PromQL query. 

Table~\ref{tab:rq3} shows the evaluation results of the contribution of the metric knowledge and system component knowledge.
When metric knowledge is lacking, both MetricAcc and QueryAcc decrease significantly, with MetricAcc decreasing at most from 91.3\% to 38.7\% and QueryAcc decreasing at most from 69.1\% to 27.8\%.
When system component knowledge is lacking, QuerayAcc also decreases significantly, at most from 69.1\% to 12.2\%.
We can see that lack of metric knowledge or system component knowledge both result in a significant degradation in the performance of \app.
The performance of \app is less degraded when lacking metric knowledge compared to when lacking system component knowledge, this is because some of the metrics in the dataset are from the open-source metric exporter.
Therefore, for some questions, the LLM itself has knowledge about the metrics.
The above results illustrate that both metrics knowledge and system component knowledge are important in the text-to-PromQL task.
\app achieves accurate retrieval of both metrics knowledge and system component knowledge, which leads to the precise generation of PromQL queries.

\subsection{RQ4: Pilot Study}
We conduct a pilot user study to investigate the usefulness of \app in helping engineers write PromQL queries.
The objective of this pilot user study is to explore whether \app can help engineers write PromQL queries more quickly compared to other tools. 
Furthermore, we also want to collect real feedback from engineers about the usefulness and other aspects of \app.

\subsubsection{Study Design}
The details of our study design are as follows.

\paragraph{Participants}
To conduct this pilot user study, we invited experienced engineers from the industry to participate. 
Our participant selection criteria required engineers to have at least five years of programming experience, be familiar with Prometheus and PromQL, have used PromQL for at least one year, and currently be responsible for developing or maintaining cloud-native online service systems.
Finally, we invite 8 engineers meeting these criteria from 5 companies in the industry, the details of the participants are shown in Table~\ref{tab:participants}. 
These participants have 5-20 years programming experience and 1-8 years PromQL writing experience.
We conduct a pre-experiment survey on their programming and PromQL writing experience, and divide them into two roughly equal participant groups ($G_1$ and $G_2$) based on the survey.

\begin{table}[]
\caption{Details of the Participants}	\label{tab:participants}
\begin{tabular}{c|c|c|cl}
\cline{1-4}
Participant & Company & \begin{tabular}[c]{@{}c@{}}Programming  Experience\end{tabular} & \begin{tabular}[c]{@{}c@{}}PromQL Writing  Experience\end{tabular} &  \\ \cline{1-4}
P1                                  & C1                              & 8 years                                                                                    & 2 years                                                                                     &  \\ 
P2                                  & C2                              & 8 years                                                                                    & 4 years                                                                                      &  \\ 
P3                                  & C3                              & 6 years                                                                                    & 2 years                                                                                      &  \\ 
P4                                  & C3                              & 10 years                                                                                   & 5 years                                                                                      &  \\ 
P5                                  & C4                              & 9 years                                                                                    & 1 years                                                                                      &  \\ 
P6                                  & C5                              & 5 years                                                                                    & 1 years                                                                                      &  \\ 
P7                                  & C5                              & 20 years                                                                                   & 8 years                                                                                      &  \\ 
P8                                  & C2                              & 6 years                                                                                    & 2 years                                                                                      &  \\ \cline{1-4}
\end{tabular}
\end{table}

\paragraph{Tasks}
We select 10 questions from the benchmark dataset for the user study.
Specifically, we randomly select 6 successful cases and 4 failed cases based on the result when using GPT-4-Turbo as the backbone LLM.
Then we divided them into two equal groups ($S_1$ and $S_2$), each containing 5 questions (3 successful cases and 2 failed cases).

\paragraph{Procedure}
Each participant is asked to write PromQL queries to solve the given questions with the aid of \app or with any other tools (such as the baseline approaches, ChatGPT, GitHub Copilot, and Search Engine).
The group of participants $G_1$ is assigned to solve the questions in $S_1$ with \app and to solve the questions in $S_2$ with any other tools.
The group of participants $G_2$ is assigned to solve the questions in $S_2$ with \app and to solve the questions in $S_1$ with any other tools.
To help participants get familiar with the monitored system, monitored metrics, and related tools.
We first created a document that introduces the system architecture, monitored metrics, and metric labels, as well as an introduction of \app and examples of its use.
Moreover, we deployed the TrainTicket system, \app, Prometheus, and the knowledge graph (stored in a neo4j database) in a Kubernetes cluster.
Three days before the experiment began, the document and the deployed platforms were provided to the participants, and we introduced the experiment process and demonstrated several example tasks to them.
During these three days, participants could familiarize themselves with the system and tools by reading the documentation, operating the system, and writing PromQL queries.
In the formal experiment, we provide participants with the original questions for each task. 
A task is considered complete only when the participant correctly queries the result, and we record the completion time. 
In addition, each question has a time limit of 10 minutes, and tasks that are not completed by the deadline are recorded as taking 10 minutes.

After all tasks are completed, we further conduct a survey to collect feedback from the participants.
Following the recommendations and practices of existing researches~\cite{survey_se,KG4CraSolver}, we selected a 4-point Likert scale to collect user feedback.
Specifically, we ask participants to evaluate the accuracy and usefulness of \app n a 4-points Likert scale~\cite{score_attitudes} (i.e., 1-disagree; 2-somewhat disagree; 3-somewhat agree; 4-agree).

\begin{itemize}
    \item Correctness: \app can generate PromQL queries correctly.
    \item Usefulness: \app is useful in helping engineers write PromQL queries.
\end{itemize}

\subsubsection{Results}

\begin{table}[t]
    \centering
	\caption{Results of the User Study}	\label{tab:rq4-user-study}
  
\begin{subtable}[t]{0.48\textwidth}
  \centering
  \caption{Average Completion Time for Each Question} \label{tab:rq4-time}
\begin{tabular}{c|c}
\hline
Tool & Avg Completion Time  \\ \hline
\app                    & $101.65$s  $\pm 22.16$                  \\
Other Tools                 & $382.63$s $\pm 120.65$                     \\
\hline
\end{tabular}
\end{subtable}
\hfill
\begin{subtable}[t]{0.48\textwidth}
  \centering
  \caption{Average Feedback Score} \label{tab:rq4-score}
\begin{tabular}{c|c}
\hline
Metric & Avg Score  \\ \hline
Correctness                    & $3.5 \pm 0.53$                    \\
Usefulness                 & $3.75 \pm 0.46$                     \\
\hline
\end{tabular}
\end{subtable}

\end{table}

Table~\ref{tab:rq4-time} shows the average completion time for each question in the user study when using \app or other tools.
It can be seen that participants can complete PromQL queries more quickly when using \app compared to other tools.
Specifically, when using \app, the average completion time for each question is only 101.65 seconds, which is only 26.6\% of the average completion time for each question when using other tools.
Table~\ref{tab:rq4-time} also presents the standard deviation for the two average completion times.
It can be seen that the completion time using \app is significantly shorter than that using other tools.
In addition, to determine whether there were significant differences in the data collected between the two participant groups, we separately conducted a t-test on the task completion time when using \app or other tools. 
Specifically, we collected the task completion time when using the same tool from the two participant groups, then tested whether significant differences existed between them.
The test results are as follows: when $\alpha$ is 0.05, the p-value for task completion time using other tools is 0.45, and the p-value for task completion time using \app is 0.97, both of which were greater than $\alpha$.
This indicates that there were no significant differences in task completion times between the two groups when using the same tools, thereby confirming that the participant grouping and task assignment did not introduce bias into our experiment.
Building upon this validated experimental design, the lower average completion time of \app compared to other tools (as shown in Table~\ref{tab:rq4-time}) demonstrates that \app can help engineers write PromQL queries more efficiently.

In the experiment, we observed that most participants spent a significant amount of time searching for the correct metrics, metric labels, and components. 
\app can automatically retrieve this knowledge and generate the corresponding PromQL queries.
Even when the generated query is not entirely correct, we observe that participants can quickly obtain the correct results after observing and modifying the output of \app.

Table~\ref{tab:rq4-score} shows the average scores rated by the participants for the accuracy and usefulness of \app.
Participants rate the correctness and usefulness with an average score of 3.5 and 3.75, respectively.
Specifically, 6 and 4 participants rated usefulness and correctness as 4, respectively.
And no participant rated any item as 2 or 1, which indicates that all participants held a positive attitude toward our tool.

During interactions with participants, they also provided some meaningful feedback.
Most of the participants mentioned that \app automatically retrieves information and generates PromQL queries are useful to improve the operational efficiency in industry systems, particularly in large-scale systems where engineers struggle to grasp the full picture of the system.
For incorrect cases, participants mentioned that it still requires time to retrieve correct information and make corrections, if some separate knowledge retrieval functions could be provided, it would help engineers quickly correct the incorrect cases.
Furthermore, some participants suggested that the tool should be able to generate multiple candidate PromQL queries, especially when the input question is broad or ambiguous.
Some participants recommended that the tool ideally should be able to automatically execute queries and fix incorrect PromQL queries, while also providing functionality to visualize metric analysis results.

In summary, the user study demonstrates \app can help engineers write PromQL more efficiently, and the participants consider \app is helpful in real systems.


\section{Discussion}

\subsection{Case Study}
To understand the capabilities and limitations of \app, we further manually analyze a successful case and a failed case in this section.
The two cases are shown in Figure~\ref{fig:cases}, where the blue part is the relevant knowledge retrieved by \app.
And all of these cases are obtained based on DeepSeek-Coder-V2-236B.

\begin{figure}[t]
	\centering
    \begin{subfigure}[t]{0.95\textwidth}
        \centering
        \includegraphics[width=0.75\linewidth ]{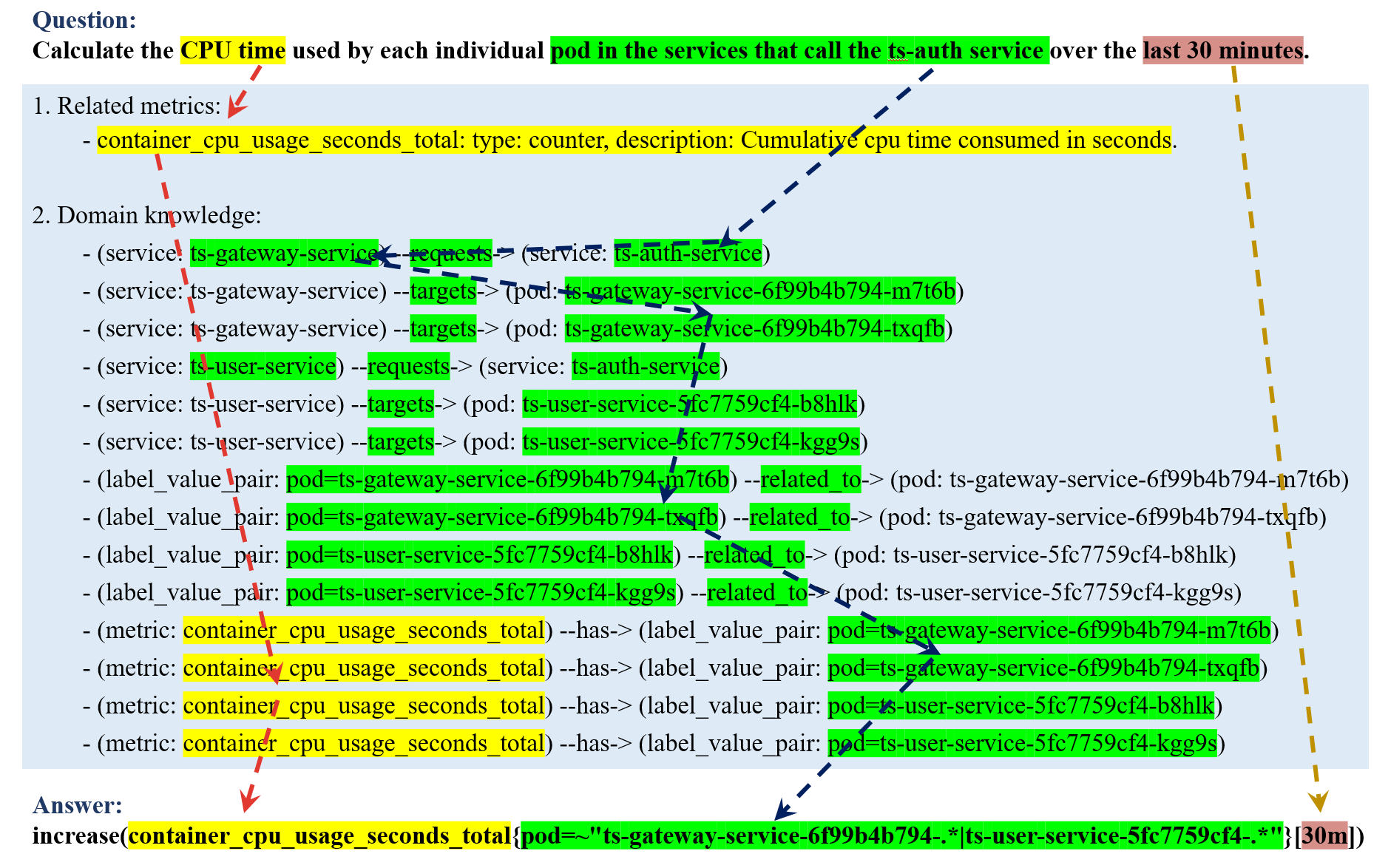}
        \caption{A Successful Case}
        \label{success_case}
    \end{subfigure}
    \begin{subfigure}[t]{0.95\textwidth}
        \centering
        \includegraphics[width=0.75\linewidth ]{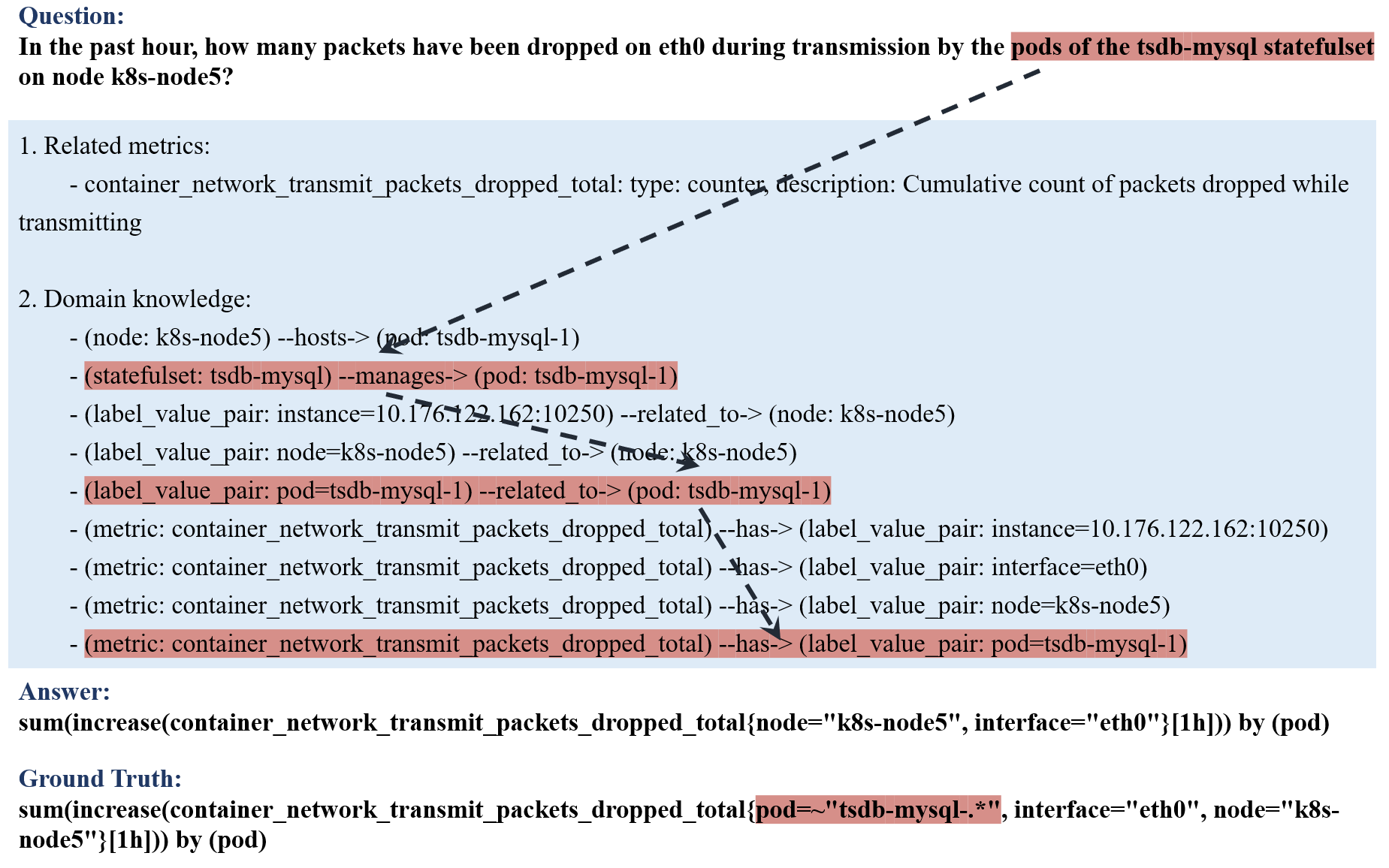}
        \caption{A Failed Case}
        \label{bad_case}
    \end{subfigure}
	\caption{Two Examples of the PromQL Query Generation}
        \label{fig:cases}
\end{figure}

\subsubsection{Successful Case}
In Figure~\ref{success_case}, we present a successful case where \app generates a PromQL query that matches the ground truth.
In this case, the engineer wants to query the CPU time of the pod of the services that call the ts-auth service.
When \app receives the question, it performs system component knowledge retrieval and metrics knowledge retrieval and then generates the PromQL query.
As shown in the green part of the figure, \app first finds the two services $ts \text{-} gateway \text{-} service$ and $ts \text{-} user \text{-} service$ that call $ts \text{-} auth \text{-} service$ and then finds the four pods corresponding to these two services.
As shown in the yellow part of the figure, \app found the corresponding metric $container\_cpu\_usage\_seconds\_total$ based on the name of the metric and the description of the metric, and then further found the related label-value pairs.
Based on the above knowledge, \app successfully generated the correct PromQL query.

In summary, \app can effectively retrieve the knowledge related to the question and utilize the capability of LLMs to generate the correct PromQL query.

\subsubsection{Failed Case}
In Figure~\ref{bad_case}, we present a failed case where \app fails to generate the correct PromQL query.
In this case, the answer output by \app is missing the label of the pod (red segment in the ground truth) compared to the ground truth.
However, we found that the labels of the pods relevant to the question have been retrieved by \app, which is the red part of the knowledge in Figure~\ref{bad_case}.
It shows that the retrieval method proposed by \app is effective, but the randomness or limitations of the capabilities of LLMs can lead to incorrect results.
On the other hand, although the output answer does not exactly match the ground truth, the query's execution result already contains the result that the engineer wants to know.
It can be seen that the answer does not use pod-related labels to further filter the output, which results in the result containing all pods on that node, whereas the pods of $tsdb \text{-} mysql$ that the user is interested in are also included.
We can consider that the answer already satisfies the engineer's needs.
Moreover, for this kind of failed case, we can fix it quickly by adding some clarification or feedback mechanism.

In summary, although sometimes the answer output by \app is not precise, it has been able to fulfill the engineer's needs. 
Meanwhile, engineers can quickly modify the output PromQL query to the correct query by referring to the knowledge retrieved by \app.
It has already helped engineers improve their working efficiency.
And in practice, we can use some clarification or feedback mechanisms to improve the generation accuracy.

\subsubsection{Failure Cause Analysis}

\begin{table}[t]
	\caption{Failure Cause Analysis in the Results of DeepSeek-Coder-V2}	\label{tab:error_cause}
    \centering
    \begin{tabular}{c|c}
    \hline
    Failure Cause                                & Number \\ \hline
    Components Relation Paths Extraction Error & 6      \\ \hline
    Metric-Component Pairs Extraction Error    & 7      \\ \hline
    System Component Knowledge Retrieval Error & 5      \\ \hline
    Metric Knowledge Retrieval Error           & 13     \\ \hline
    Query Generation Error                     & 45     \\ \hline
    \end{tabular}
\end{table}

To understand the limitation of \app, we manually analyze the failed cases in the results of DeepSeek-Coder-V2.
We classify the causes of failure according to the steps of \app, Table~\ref{tab:error_cause} summarizes the causes and distributions of the failed cases.

Among all failed cases, most failures are occurred in the final query generation step, with a total of 45 cases (59.21\%).
In these cases, \app correctly retrieves all the necessary knowledge, but the PromQL query generated by LLM is incorrect.
After manual analysis, we find the causes of these errors included: incorrect use of labels, incorrect use of functions, calculation logic errors, syntax errors, etc.
Among them, incorrect use of labels, similar to the case shown in Figure~\ref{bad_case}, is the most common error (22 cases).
Meanwhile, incorrect use of functions and calculation logic errors each had 8 cases.
We think that these generation errors are primarily due to the fact that the general-purpose code LLM has not been fine-tuned for PromQL.
The LLM lacks a comprehensive understanding of labels, functions, calculation logic, and other elements in PromQL.
In the future, we can address these issues by training an LLM specific to PromQL and using it for PromQL query generation.

In the remaining cases, apart from a relatively high number of metric knowledge retrieval errors, other errors are generally distributed evenly.
Errors in retrieving metric knowledge and system component knowledge are mainly caused by similar metrics, labels, or component names.
Component relation paths extraction errors are mainly caused by the LLM extracting non-existent relations or entities.
Metric-Component pairs extraction errors are mainly caused by the extracting metrics that do not exist in the extracted system components.
While these errors occur in the evaluation, they only account for 13.48\% of all test cases.
This indicates that the question parsing and knowledge retrieval methods proposed by \app are effective.
In the future, we can use fine-tuned models or provide more knowledge during question parsing to avoid extracting non-existent metrics, components, or relationships. We can also design more comprehensive retrieval methods to improve retrieval accuracy.

In summary, we analyze the causes of the failure cases and find that most errors occurred during the query generation step.
In the future, we can train specific LLM for PromQL to improve the accuracy of the generation.




\subsection{Threats to Validity}

\subsubsection{Internal Validity}
The threat to internal validity mainly lies in the implementation and configuration of baseline approaches.
The basic prompt is simple to implement and no configuration is involved.
There is no directly available open-source implementation of the few-shot learning-based approach.
Therefore, we implement the approach by referring to the example of question answering using embeddings-based search provided in the OpenAI cookbook~\cite{openai_cookbook}.
We carefully check the implementation and choose the optimal configuration.

\subsubsection{External Validity}
The threat to external validity mainly lies in the generalizability of our approach and the representativeness of the dataset.

The generalizability of our approach is under threat of dependency on Kubernetes and Prometheus.
On the one hand, the knowledge graph schema in this paper is designed for Kubernetes-based cloud native online service systems and may not be suitable for all systems.
This makes \app difficult to use directly in other types of system.
On the other hand, the prompts, cots, and examples in \app are all designed for Prometheus's data model and PromQL.
This makes it difficult to directly apply \app to other metric monitoring platforms.
To integrate PromCopilot with other monitoring platforms, the prompt, cots, and examples need to be modified according to the platform's data model and DSL.
If the data model is significantly different from Prometheus's, further modifications may be required to entities and relationships related to metrics in the knowledge graph, as well as the logic for metric knowledge retrieval.


Our experiments are conducted on the dataset we constructed.
Because no publicly available dataset fits our task, we constructed the first text-to-PromQL dataset.
The dataset are constructed based on the open-source microservice system TrainTicket, as there are no publicly available datasets from industrial.
It is one of the largest open-source microservice systems and have been widely used in existing studies~\cite{deeptralog,nezha, microeval,eadro}.
It contains 96 service instances after deployment, which is close to some medium-sized microservice systems.
Therefore, it is possible to simulate real metrics query scenarios.
We don't use other open-source systems such as Sock Shop, because their scale is too small to model complex metrics query scenarios.
Moreover, the questions in our dataset are carefully designed.
When constructing the dataset we refer to official documents, questions on StackOverflow, and real cases from Alibaba Group.
Additionally, we have invited engineers to help us confirm the soundness of the questions.
Therefore, the questions in our dataset are representative and can be used to evaluate the effectiveness of the text-to-PromQL approaches.
Since we only referenced common metric query scenarios and the authors' experience is limited, our dataset may not cover some query tasks.
Additionally, a single metric query question may correspond to multiple PromQL queries. Although we have provided several answers for these questions, we may still not have covered all possible queries.

\subsubsection{Construct Validity}
The threat to construct validity mainly lies in the choices of evaluation metrics in the experiment.
Currently, there are no evaluation metrics specifically designed for PromQL generation tasks.
Therefore, based on the characteristics of PromQL, we have designed three evaluation metrics: MetricAcc, SyntaxAcc, and QueryAcc. These metrics assess the effectiveness of PromQL generation from three perspectives: metrics, syntax, and query statements.
Although these three aspects cover key components of PromQL queries, some important aspects may still be overlooked.
We will explore more effective evaluation metrics in the future.

\subsubsection{Conclusion Validity}
The threat to conclusion validity mainly lies in the possible bias of user study results.
Our study involved only eight participants, each completing just ten tasks, which may result in insufficient statistical significance.
To mitigate this risk, all invited participants have at least five years of programming experience and have used PromQL for over a year.
In addition, we divided participants into two groups for the study.
Significance tests on the results from the two groups confirmed no significant differences between them.
These measures ensure our findings have some representation. 
We will explore to conduct larger-scale user studies in the future.

\section{Related Work}
\label{section:relatedWork}

\textbf{Natural Language to Domain-Specific Language}:
In recent years, deep learning and machine learning techniques have been widely used in text-to-DSL tasks, such as text-to-SQL~\cite{text-to-sql-survey1, text-to-sql-survey2} and text-to-GraphQL~\cite{text-to-graphql-ISF}.
These studies want to enable users to interact with domain-specific tools or data through natural language.
Text-to-SQL has become one of the most popular research topics, as SQL is widely used for data querying and management.
With the development of LLMs, there have been several studies in recent years exploring using LLMs to implement text-to-SQL~\cite{llm-text2sql-emprical1,llm-text2sql-emprical2,llm-text2sql-emprical3}.
Liu et al.~\cite{zero-shot-text-to-sql} evaluate the text-to-SQL capability of ChatGPT under zero-shot setting, demonstrates the potential of LLMs for the task.
Dong et al.~\cite{c3} propose a ChatGPT-based zero-shot text-to-SQL apprach, which outperforms the performance of fine-tuning-based approaches through clear prompting, calibration with hints and consistent output.
Sun et al.~\cite{sql-palm} enhance text-to-SQL performance of LLMs based on few-shot learning and instruction tuning.
To the best of our knowledge, there is no research focusing on text-to-PromQL.

\textbf{LLMs for AIOps}
To date, many efforts have been dedicated to LLM-based AIOps, including AIOps Q\&A~\cite{micro_qa,opsEval}, root cause localization~\cite{few-shot-learning-rca,llm-rca-icse23,RCACopilot}, log parsing~\cite{divlog,lilac}, and incident mitigation~\cite{Nissist}.
Liu et al.~\cite{opsEval} evaluate the capabilities of LLMs on AIOps Q\&A tasks and discuss several potential research directions.
Quevedo et al.~\cite{micro_qa} implement question-answering for microservice architectures using large models and static code analysis.
Ahmed et al.~\cite{llm-rca-icse23} uses historical fault records to fine-tune the LLMs for root cause localization and mitigation steps recommendation.
Chen et al.~\cite{RCACopilot} uses multi-source diagnostic information to model historical faults, and achieve root cause and fault category prediction based on few-shot learning.
Zhang et al.~\cite{few-shot-learning-rca} uses a similar idea to achieve automated root cause localization based on GPT-4.
And Jiang et al.~\cite{lilac} and Xu et al.~\cite{divlog} use few-shot learning to achieve LLM-based log parsing results in promising performance.
An et al.~\cite{Nissist} implement automated incident mitigation based on the troubleshooting guides summarized by engineers by leveraging LLMs' ability to use tools and execute code.
However, above works are based on historical fault cases and it is difficult to achieve good results when dealing with new faults.
Some studies explore using LLMs to assist operation engineers in performing some operation tasks.
Yu et al.~\cite{monitorassistant} use LLMs to assist engineers in configuring metric monitor rules.
Jiang et al.~\cite{xpert} used few-shot learning to recommend fault-related metric query statements for operation engineers.
These studies are more practical than others, but they are also limited by historical data, which makes it difficult to achieve good generalizability.
In this paper, we propose a text-to-PromQL approach that does not rely on historical data, which can help engineers query metrics in a simple and fast way.

\section{Conclusion}
In this paper, we proposed \app, an LLM-based text-to-PromQL framework that enables engineers to interact with Prometheus metrics data through natural language.
\app uses a knowledge graph to describe the complex context (e.g., metric names, metric labels, system component dependencies) of an online service system.
Then, \app transforms the engineers' natural language questions into PromQL queries through the synergistic reasoning of knowledge graph and LLMs.
To evaluate \app, we manually construct the first text-to-PromQL benchmark dataset based on an open-source microservice system.
The experiment results demonstrate the effectiveness of \app in text-to-PromQL.
When using GPT-4-Turbo as the backbone LLM, PromCopilot achieves an accuracy of 69.1\% in generating PromQL queries.
Moreover, the results confirm that \app is accurate in retrieving metric knowledge and system component knowledge.

\app currently targets cloud-native online service systems and PromQL.
And we think \app has the potential to extend to other types of systems and metric query languages.
In the future, we will explore applying \app to application systems with more diverse architectures and different metric monitoring platforms.
We will try to design specialized system context knowledge graphs for application systems with different architectures and design knowledge retrieval methods and prompts for different metric monitoring systems.
Furthermore, we will explore improving the PromQL generation capabilities of generic LLMs through techniques such as fine-tuning.



\section{Data Availability}
All the data and code of the work can be found in our replication package~\cite{PromCopilot}.

\begin{acks}
This work was supported by the Natural Science Basic Research Program of Shaanxi (Program No.2025JC-YBQN-851).
\end{acks}

\bibliographystyle{ACM-Reference-Format}
\bibliography{reference}


\end{document}